**Blocking of the CD80/86 axis as a therapeutic approach to prevent progression to more severe forms of COVID-19**

Antonio Julià[1,*], Irene Bonafonte[1], Antonio Gómez[1], María López-Lasanta[1], Mireia López-Corbeto[1], Sergio H. Martínez-Mateu[1], Jordi Lladós[1], Iván Rodríguez-Nunez[2], Richard M. Myers[2], Sara Marsal[1,*]

[1]Rheumatology Department and Rheumatology Research Group, Vall d'Hebron Hospital Research Institute, Barcelona, Spain.

[2]HudsonAlpha Institute for Biotechnology, Huntsville, Alabama, USA.

*Corresponding author: toni.julia@vhir.org (AJ); sara.marsal@vhir.org (SM)

## Abstract

In its more severe forms, COVID-19 progresses towards an excessive immune response, leading to the systemic overexpression of proinflammatory cytokines like IL6, mostly from the infected lungs. This cytokine storm can cause multiple organ damage and death. Consequently, there is a pressing need to identify therapies to treat and prevent severe symptoms during COVID-19. Based on previous clinical evidence, we hypothesized that inhibiting T cell co-stimulation by blocking CD80/86 could be an effective therapeutic strategy against progression to severe proinflammatory states. To support this hypothesis, we performed an analysis integrating blood transcriptional data we generated from rheumatoid arthritis patients treated with abatacept -a CD80/86 co-stimulation inhibitor- with the pathological features associated with COVID-19, particularly in its more severe forms. We have found that many of the biological processes that have been consistently associated with COVID-19 pathology are reversed by CD80/86 co-stimulation inhibition, including the downregulation of IL6 production. Also, analysis of previous transcriptional data from blood of SARS-CoV-infected patients showed that the response to abatacept has a very high level of antagonism to that elicited by COVID-19. Finally, analyzing a recent single cell RNA-seq dataset from bronchoalveolar lavage fluid cells from COVID-19 patients, we found a significant correlation along the main elements of the C80/86 axis: CD86+/80+ antigen presenting cells, activated CD4+ T cells and IL6 production. Our in-silico study provides additional support to the hypothesis that blocking of the CD80/CD86 signaling axis may be protective of the excessive proinflammatory state associated with COVID-19 in the lungs.

## Introduction

Infection with SARS-CoV2 can lead to different degrees of symptomatology and severity, ranging from asymptomatic to an extreme immune response leading to patient death (1). The fatality rate of COVID-19 is estimated to be close to 1%, which is 10 times more than typical seasonal influenza (2). COVID-19 has been associated to the cytokine storm or cytokine release syndrome (CRS) (3), an overload of proinflammatory cytokines that leads to massive organ failure, predominantly uncontrollable lung inflammation that, even with the help of mechanical ventilation, can lead to organ collapse and death (4). Its high capacity of dissemination and the severity stages at which it can lead, have contributed to one of the largest pandemics since the Spanish flu in 1912 (5). There is currently a major international effort to develop vaccines against SARS-CoV2 (6). However, vaccine development is a lengthy process that needs to ensure both effective virus neutralization and patient safety(7). Therefore, while vaccines are being developed, there is a need to identify therapies that can help reduce the symptomatology associated with COVID-19 and, mostly, prevent infected individuals from progressing into life-threatening stages(8).

Many different strategies are being contemplated to find drugs to treat COVID-19. These include in-silico docking strategies (9), protein-protein network interactions (10)(11), in vitro testing (12), and also evaluation of disease progression to severe forms in patients undergoing different therapies. In inflammatory rheumatic disease, many patients are treated with drugs targeting specific components of the immune system (e.g. Tumor Necrosis Factor-alpha (TNF) and Interleukin-6 (IL6) signaling, and CD80/86 T cell co-stimulation) to dampen the excess immune activation that characterizes these diseases. Closely evaluating these patients can provide invaluable information on how blocking certain elements of the immune response can be beneficial or detrimental towards COVID-19 severity. Using a cohort of 959 rheumatic patients under different targeted therapies, we have recently provided epidemiological evidence (13) that suggests that anti-IL6 receptor therapy (tocilizumab) and CD80/86 blockade with CTLA4-Ig (abatacept) are associated with a lower prevalence of COVID-19 associated symptomatology as defined by the world health organization. Given that severe COVID-19 patients express high levels of IL6 (14), targeting the signaling of this cytokine -either directly or through its receptor- has been contemplated and is currently being used off-label to attempt to rescue critically ill patients (15). CTLA4-Ig, however, has not been so far proposed as a therapeutic agent for COVID-19 severity. Abatacept is a fusion protein consisting of the extracellular domain of human cytotoxic T lymphocyte antigen 4 (CTLA4) linked to the modified Fc domain of human IgG1 (16). It binds to both costimulatory proteins CD80 and CD86 on professional antigen presenting cells -dendritic cells, macrophages and B cells- with higher affinity than CD28 on the surface of T cells, thereby preventing the necessary costimulatory signal required by T cells to progress to activation. It was originally developed to treat rheumatoid arthritis, where T cell activation is central to the disease pathology, but has currently been approved also for psoriatic arthritis and juvenile idiopathic arthritis (16).

In severe COVID-19, macrophages in the lung activate and produce large amounts of IL6. Single cell analysis on bronchoalveolar lavage fluid (BALF) from COVID-19 patients suggest that monocyte-derived FCN1+ macrophages substitute alveolar macrophages in the lung during inflammation (17). In either case, how macrophages drastically increase the proinflammatory levels in COVID-19 is yet not known(8). High levels of T-cell mediated activation could be one of the possible causes. In turn, T cells could become hyperactivated due to excessive activation through the CD80/86 axis (**Figure 1**). Therefore, dampening this activation signaling pathway could be an effective therapeutic approach to prevent immune system hyperactivation, and progress to more severe stages. To support this hypothesis raised by the findings of our previous epidemiological survey in rheumatic patients, here we performed an in-silico study using transcriptional data from abatacept-treated patients as well as from COVID-19 patients. Our data support a significant antagonism of abatacept on COVID-19-associated processes at the systemic level, and suggest that blocking the CD80/86 axis could be a useful therapeutic approach to reduce the level of macrophage-associated inflammatory response.

## Methods

### Patients and samples

A total of n=38 rheumatoid arthritis (RA) patients starting treatment with abatacept were recruited in the framework of the PACTABA project (BMS). All patients were treated with subcutaneous abatacept at the recommended weekly dosage of 125 mg. The PACTABA project is a multicentric Spanish pharmacogenomic study performed in a subset of patients from the ASCORE clinical trial (Bristol-Myers Squibb, ClinicalTrials.gov Identifier: NCT02090556). This observational and prospective study was designed to estimate the retention rate of subcutaneous abatacept over 24 months in the routine clinical practice of RA patients (18). The study was approved by the institutional review board and informed consent was obtained in all cases.

Whole blood samples were obtained at the start of the therapy with abatacept and at week 12. Blood was collected using RNA-stabilizing PaxGene tubes (PreAnalytiX, Switzerland), which preserve total RNA from the time of venipuncture. Total RNA was extracted using the PaxGene blood isolation kit (Qiagen). All samples had a RIN index > 7 and were included for RNA-seq analysis. RNA-seq libraries were performed using with the KAPA RNA HyperPrep Kit, with RiboErase (HMR) Globin (globin and rRNA depletion protocol) for Illumina sequencing platforms and with the addition of Unique Molecule Identifiers (UMIs). Sequencing was performed using the NovaSeq platform (Illumina) at average of 250 paired-end reads.

**RNA data preprocessing**

For our RA patient cohort, FASTQ files were aligned to the GRCh37 human reference genome assembly using STAR (19). After alignment, we performed deduplication of the PCR duplicates using UMItools (20) to mark and after removing with samtools (21). Gene-level read counts of the deduplicated bam files were obtained with *featureCounts* (22). The RA longitudinal dataset will be made available through the NCBI Gene Expression Omnibus database (GEO link available after data has been deposited).

To characterize the gene expression profile associated with COVID-19, the RNA-seq data from Xiong et al on peripheral blood mononuclear cells (PBMCs) from three patients infected with COVID-19 and three healthy controls was used (23). The raw sequencing data was processed as described previously. In order to infer the purity and cell composition of this dataset, we used the ABIS immune cell deconvolution method which has been specifically developed for PBMC data (24). This analysis revealed potential neutrophil presence, particularly in one sample (**Supplementary Figure 1**). To compensate for this potential confounding effect, the estimated percentage of granulocytes was used as a covariate in the differential expression analysis.

**Differential expression and gene set enrichment analysis**

Genes differentially expressed in COVID-19 patients were determined using *edgeR* (25). Raw counts were normalized by trimmed mean of mean values (TMM) normalization, and genes with low expression values (i.e. < 2 counts per million -CPM- in at least 2 samples). Differentially expressed genes were determined by fitting a quasi-likelihood negative binomial generalized linear model and adjusting for multiple testing by Benjamini & Hochberg (26). Genes with an adjusted p-value < 0.05 were considered significant.

To test for gene expression changes induced by treatment with abatacept at a systemic level, the differential expression between baseline and week 12 was tested in RA patients using *limma* (27). Raw counts were normalized using TMM normalization for all genes with > 0.6 CPM in at least 20 samples. Normalized counts were log-transformed to log-counts using *voom* (28) and a linear model was fitted blocking by individual and adding sex, age and batch (i.e. library plate) as covariates. The percentage of granulocytes in each sample were estimated using DeconCell (29), and also included as a covariate in the model in order to make the results more directly comparable with the results obtained with the COVID-19 dataset.

Biological processes affected either by SARS-CoV-2 infection or treatment with abatacept were determined using the Gene Set Enrichment Analysis (GSEA)(30) implemented in the *fGSEA* R package. To rank the genes based on their association to COVID-19 and abatacept, the minus log of the p-value of the differential expression analysis was calculated and multiplied by the sign of the log fold change as described previously (31). We included the sign of the log fold change in order to preserve the directionality of the enrichment and, therefore, to be able to determine if the biological process was either activated or repressed by the viral infection and by the treatment with abatacept. Biological processes with a FDR < 0.05 and a normalized enrichment score (NES) > 0 were considered to be up regulated, while biological processes with a FDR < 0.05 and a NES < 0 were considered to be down regulated.

**Analysis of the antagonism of the abatacept signature to COVID-19 associated processes**

To identify COVID-19 biological processes we used two alternative strategies. In the first, we identified a list of biological processes that have been associated to COVID-19 pathology through diverse clinical and biological studies. In particular, biological processes that have been associated with COVID-19 severity were prioritized. Those pathological processes that were unlikely to be captured at the blood transcriptional level (e.g. lung fibrosis) were excluded. A total of 22 processes were finally selected, which are divided in a group of 6 processes related to viral immune sensing and anti-viral response (i.e. first stage of COVID-19)(5), and a group of 16 processes related to hyperinflammation and severity (i.e. second stage)(32)(3)(33)(34). **Table 1** describes the selected processes and the most relevant bibliographical evidence. To evaluate their modulation by abatacept, we selected a representative gene set from the Biological Process (BP) database from the Gene Ontology (GO)(35) and tested their association to treatment with GSEA as described previously.

In the second approach, we used the available PBMC transcriptional data generated from COVID-19 patients and controls (14) to determine the biological processes associated with COVID-19. Similarly, biological processes regulated by abatacept at the systemic level were determined by comparing the baseline gene expression of RA patients to the expression after 12 weeks of treatment with abatacept. To avoid testing BPs represented by very low gene numbers or excessive gene content (too broad process annotation), we selected only GO terms with > 10 genes and < 300 genes. To account for the large number of GO terms tested, a false-discovery rate (FDR) adjustment was performed. GO terms with FDR < 0.05 were considered significant. Since many GO terms can be very similar in their gene composition, we reduced the redundancy in the analyzed gene sets using a distance measure based on the Jaccard index and hierarchical clustering. Within each cluster, the GO term showing the highest association to COVID was chosen to represent the corresponding biological process.

**Single cell RNA-seq analysis of COVID-19 BALF cells**

In order to evaluate the potential utility of CTLA4-Ig blockade of CD80/86 T cell signaling in COVID-19-mediated lung inflammation, we analyzed single cell RNA-seq data from bronchoalveolar lavage fluid (BALF) cells. Single-cell RNA-seq raw data from BALF samples from 9 COVID-19 patients (n=6 severe and n=3 mild) and 3 controls were downloaded from the GEO database (accession GSE145926). This dataset was generated using the 10x genomics platform and includes samples used in a publication demonstrating the predominant influx of monocyte-derived macrophages in COVID-19 patients (17). The raw data processing and analysis -including, normalization, scaling, and clustering of cells- was performed using Seurat (v3)(36)(37) and scTransform (38). Cells with <200 or >6,000 unique feature counts, >10% mitochondrial counts or < 1,000 UMIs were filtered out. Samples were log-normalized and scaled for the number of genes, number of UMIs and percentage of mitochondrial reads. Cell type clustering was performed using the “FindClusters” function from Seurat. In brief, this method uses a shared nearest neighbour (SNN) modularity optimization-based clustering algorithm to identify clusters of cells based on their PCs. Before constructing the SNN graph, this function calculates the k-nearest neighbours. The number of PCs used for each clustering round (k=50) was estimated by the evaluation of the elbow of the PCA scree plot.

## Results

### The transcriptional changes induced by CD80/86 blocking antagonize COVID-19 associated processes

In our first approach, we evaluated how COVID-19 pathological processes were affected by treatment with CTLA4-Ig at the transcriptional level. Of the 22 curated processes, we found that abatacept induced transcriptional changes in 16 of them (72.7%, binomial test P=8.5e-17). All significant changes occurred in the opposite direction to that described by previous clinical and experimental studies on COVID-19 **Figures 2A and 2B** show the observed changes induced by CTLA4-Ig in the anti-viral primary response and hyperinflammation biological processes, respectively. **Figures 3A and 3B** show a detailed visualization of the differential gene expression for the selected pathological processes representing anti-viral immune response and hyperinflammation-mediated severity, respectively.

In our second approach, we identified a total of 260 pathways differentially activated in COVID-19 patients compared to controls (FDR < 0.05) (**Supplementary Table 1**). These pathways were mostly associated to phagocytosis, endocytosis and lysosome function, immunoglobulin mediated immunity, antigen processing and presentation, acute inflammatory response and cytokine cascades (i.e. IL1, IL6 and IL8 production, TNF signaling, type I and type II interferon signaling, NF-kB signaling and macrophage activation), metabolic processes, mitochondrial activity, cell cycle, response to reactive oxygen species and apoptosis. Other interesting pathways included the down-regulation of T cell receptor signaling, the over-activation of the cellular response to angiotensin, the up regulation of transferrin transport and the regulation of blood pressure.

In the RA patient cohort, comparing week 12 to baseline gene expression profiles, we found 109 pathways associated with CTLA4-Ig therapy (FDR < 0.05, **Supplementary Table 2**). These pathways were associated predominantly to cell cycle and cell division, myeloid cell differentiation, antigen processing and presentation, immunoglobulin mediated immunity, acute inflammatory response, cytokine cascades (IL7 and IL6), phagocytosis, endocytosis and lysosome function, ribosomal RNA related processes and apoptosis. Comparing the two datasets, we found a total of 49 overlapping significant pathways (**Figure 4**). From these, 47 (96%) were found to be antagonistically activated by abatacept compared to COVID-19. Among the significant antagonistic processes, we found 16 pathways related to immune system processes including processes related to viral defense (viral transcription, GO:0019083), innate immune system activation (Fc receptor signaling pathway, GO:0038093; myeloid leukocyte differentiation: GO:0002573) and acquired immune response (regulation of B cell activation, GO:0050864; humoral immune response mediated by circulating immunoglobulin, GO:0002455) and cytokine production (interleukin-6 production; GO:0032635). Other abundant significant processes included those related with cell cycle and RNA transcription (n=18) and with endocytosis (n=5). Only one pathway (mitochondrial translational elongation, GO:0070125) was activated and one process (mucosal immune response: GO:0002385) was inactivated by both exposures. The probability that the observed antagonism occurred by chance, is very low (P value < 2.5e-18, binomial test). **Figure 5** shows the most significant processes antagonized by CD80/86 blocking with abatacept. The complete list of overlapping -antagonistic and agonistic- pathways is included in **Supplementary Table 3**.

**scRNA-seq analysis of CD80/86 axis in BALF**

In order to evaluate the relevance of CD80/86 co-stimulation in COVID-19 pathology we analyzed the correlation of three key cell types participating in this axis: CD80+ or CD86+ antigen-presenting cells, activated CD4+ T cells and IL6-producing macrophages. For this objective we analyzed a recent scRNA-seq dataset generated from 12 BALF samples (17).This dataset consists on samples from six severe and three moderate COVID-19 patients, as well as a set of CD45+ selected BALF cells from healthy controls. In total, 54,420 cells passed our quality control analysis and were analyzed using the scTransform algorithm (38). With this approach a total of 34 cell clusters were identified (**Figure 6**).

Most of the cell clusters aggregated into four major regions representing i) T and NK cells (8,014 cells, 14.7%, clusters 9, 11, 5, 25 and 18), ii) two regions of FABP4+ alveolar macrophage cells (17,399 cells, 32%, clusters 1, 2, 3, 30, 7, 32 and 12), and iii) a large cluster aggregate of active, IL6-expressing macrophages both from FABP4+ alveolar and monocyte-derived FCN1+ macrophages (24,742 cells, 45.5%, clusters 6, 20, 13, 22, 0, 8, 4, 15, 10, 16, 14, 26). Both alveolar macrophage-only clusters belong to the three healthy controls (**Supplementary Figure 2**). As described by the original study(17), there is a marked transition from FABP4+ alveolar macrophages to FCN1+ monocyte-derived macrophages from healthy to infected, and as COVID-19 severity increases (**Figure 7 A**).

When looking at the co-stimulation proteins that are the target of CTLA4-Ig -CD80 and CD86- we found that healthy controls had a significantly lower expression of both proteins (**Figure 7 B**). This difference was high for CD86 (26.5% CD86+ in controls vs 45.4% in COVID-19 infected patients, P<1e-16) and drastically different for CD80+ cells (1.7% CD86+ in controls vs 19.6% in COVID-19 infected patients, P<1e-16). Next, we identified the set of cells showing production of IL6, the key cytokine in COVID-19 severity and the cytokine storm. Among the different BALF cell clusters, we found IL6-expressing cells almost exclusively in the activated macrophage cluster derived from COVID19 patients (**Figure 7 C**). Finally, we identified the cluster containing the CD4+ T cell element of the CD80/86 axis. Cluster 5 was found to aggregate markers of different types of CD4+ T cells including CCR7 (naïve TCD4), IL2RA (Treg), FOXP3 (Treg), IL7R (naive), LTB (naive), CXCL13(T peripheral helper) (**Figure 7D**). This cell cluster also was characterized by the expression of CTLA4, the protein that is recombined in abatacept and also a marker that is highly upregulated in active T cells and Treg cells (39), and was therefore used to represent the CD80/CD86 activated T CD4+ element.
We next tested for association between the simultaneous presence of each element, of the CD80/86 axis. First, we found a significant correlation between CD80+ or CD86+ APCs and active CD4+ T cells ($r^2$=0.87, P=0.00036; and $r^2$=0.73, P=0.0069) (**Figure 8A and Figure 8B**). Second, we also found evidence of correlation ($r^2$=0.79, P=0.0024) between the number of activated CD4+ T cells and the number of IL6 producing cells (**Figure 8C**). Finally, we found a significant correlation between the number of CD80+ or CD86+ cells and IL6+ production ($r^2$=0.93, P=1.5e-7, and $r^2$=0.82, P=0.0011, respectively) (**Figure 8D**).

## Discussion

Since the beginning of the pandemic, SarsCov-2 has infected >4 million people and caused at least 280,000 deaths world-wide as of May 11th 2020 (https://coronavirus.jhu.edu/map.html). There is currently a pressing need to identify therapies that can help prevent and treat patients with more severe symptoms of COVID-19. Based on previous epidemiological evidence from rheumatic treated patients on several targeted immunomodulators, we have found evidence that CTLA4-Ig (abatacept) could be protective of COVID-19 symptomatology. Here we provide additional in-silico evidence to support the link between the relevance of the CD80/86 axis and COVID-19. We have characterized the transcriptional response of abatacept-treated RA patients and we have found that is highly antagonistic to key biological processes that have been linked to COVID-19 severity. Using gene expression data from COVID-19 infected patients, we have also found that this antagonism extends significantly to many other biological processes induced by SARS-CoV2 virus infection. Finally, analyzing single cell data from BALF samples, we have found evidence that the CD80/86 axis is closely linked to IL6 expression in the BALF macrophage compartment in COVID-19.

Our results suggest that blocking the CD80/86 axis could be useful to dampen the hyperactivation of the immune response in severe COVID-19, and subsequently reduce the damage to the lungs and the excessive systemic cytokine production. Once SARS-CoV2 enters the lung through attachment to ACE2 expressed on alveolar epithelial cells (40)(41), an immune response is subsequently mounted. Central to this response, phagocytic cells, initially alveolar macrophages that are eventually replaced by massively infiltrating monocytes (17), are highly activated, releasing a large number of cytokines. Macrophages act as powerful antigen-presenting cells (APCs) on activated CD4+ T cells to further amplify the level of immune response. For this objective, the co-stimulatory signal in the form of CD80 and CD86 expression in the surface of APCs is essential, otherwise an anergic program is initiated (42). From all organs, the lung is the tissue with the highest levels of CD80 mRNA, and the second tissue after the spleen with the highest levels of CD86 mRNA expression (**Supplementary Figure 3**). Even CTLA4 itself is highly expressed in the lung compared to the rest of tissues. This high expression of the elements of the CD80/86 axis might reflect the need of the lung to respond rapidly to potential immunological insults in a tissue that is constantly exposed to environmental cues (43). However, having this natively high potential for antigen presentation and T cell activation could be a detrimental factor in a rapidly spreading virus like SARS-CoV2. A larger cell mass capable of activating T cells would in turn lead to a more pronounced stimulation of IL6 producing programs (44). This high signal amplification potential could therefore explain the sudden transition into the extremely high cytokine production stage known as cytokine storm (45).

In our analysis approach, treatment with CTLA4-Ig showed a regulation at the transcriptional level that is highly antagonistic to that induced by COVID-19. We have shown this by two complementary strategies. In the first strategy we have found that, from the most well-established pathological processes associated with COVID-19 to date, abatacept induces a transcriptional change that is contrary to the one induced by the virus in most of them. In a first group of processes related to innate immune system detection of SARS-CoV2 we found that treatment with abatacept downregulates TLR-signaling pathway and endosomal transport. TLRs are key to detect the presence of the virus, leading to the activation of proinflammatory transcription factors and the expression of IFN and other cytokines(46). While viral detection is essential to trigger the immune response in the first stages of the infection, later its utility might be superseded by the need to adequately control excessive inflammatory response (5). To this regard, our analysis did not detect a strong type I IFN inhibition, thereby suggesting that this basic anti-viral mechanism is protected. Of interest, we found a significant downregulation of endosomal transport, a biological process that has been linked to the possible protective of hydroxychloroquine in COVID-19 (33). A reduction endosomal development and maturation would keep at stake viral load. However, the confirmation of this mechanism in relation to SARS-CoV2 infection and the clinical protection are still controversial and are in need of robust evidence (47).

Abatacept inhibits many biological pathways that are key to the hyperinflammatory stage. As described previously, the immune response to SARS-CoV2 can be divided into two phases, one first phase where the antiviral response is deployed and a second phase of excessive inflammatory response (27). The latter might not activate in many infected patients, but likely occurs in most of those who progress to more severe symptoms, particularly those with increasing lung inflammation. This stage is characterized by the activation of myeloid cells, mostly monocyte-derived macrophages that massively infiltrate the lung (8). We have found that treatment with CTLA4-Ig significantly down-regulates myeloid leukocyte differentiation and macrophage activation. This specific anti-inflammatory role could be a useful property to avoid entering into the life-threatening massive cytokine release stage. To this regard, we have found that abatacept downregulated both the production and the response to several key cytokines of the severe stage in COVID-19 including TNF (48), IL1 (33), IL8 (49), IL7 (48) and IL6 (48). Importantly, we have found that downregulation is strong for IL6, the cytokine that is more abundantly produced in patients progressing to severity both in serum and in the lungs (45). Therapies aimed at IL6 signaling, tocilizumab and sarilumab, are currently being evaluated in at least three clinical trials by 11$^{th}$ May 2020, and there is increasing reporting of beneficial effects in off-label use in severe patients (50). In line with this, our previous epidemiological study, both CD80/86 blocking and anti-IL6R therapies showed the lowest incidence of COVID-19 suspected cases among rheumatic patients (13). Consequently, the potential beneficial effect of CTLA4-Ig could be due to the upstream regulatory effect of this central cytokine for COVID-19 pathogenesis.

Abatacept also strongly downregulated two additional immune processes: complement activation and B cell mediated immunity. Complement over-activation has been identified in the lungs of COVID-19 and antibody therapies against elements of the complement system are being currently being evaluated to treat severe cases (32). The downregulation of the B cell mediated response, could also be beneficial to avoid overactivation of the myeloid compartment. While neutralizing antibodies to SARS-CoV are associated to a reduction in the viral load (51), there is also the possibility that high titers antibodies could lead to a more aggressive immunological response. During the SARS-CoV1 epidemic, neutralizing antibody levels were higher in deceased patients (52), which raised the possibility of antibody-dependent enhancement (ADE) contributing to disease exacerbation. In this situation, non-neutralizing antibodies facilitate the phagocytosis of the virus into macrophage through the Fc-receptor leading to their activation into proinflammatory phenotypes (53). ADE has been previously identified in other coronaviruses (54)(55), and experimental evidence in rhesus macaques infected with SARS-CoV1 found that anti-spike IgG antibodies contributed to massive influx of monocytes in the lung and subsequent severe acute injury (56). In this context, the effect of reducing B cell immunity by abatacept might be beneficial by reducing the likelihood of low-affinity antibodies that would contribute to ADE-induction (7). To this regard, evidence showing that antibodies raised against SARS-Cov2 can lead to ADE would be of high relevance, particularly in the development of safe vaccines.

The results of this study are entirely in the in-silico domain and therefore they must be considered as additional evidence on top of increasing evidence from the clinical domain. Consequently, additional independent data from large patient cohorts will be useful to confirm this hypothesis. For example, the Global Rheumatology Alliance (57), an international collaborative initiative that is currently aggregating data from patients around the world, will likely become an ideal resource to confirm the observed lower at-risk incidence in abatacept-treated rheumatic patients (13). Among the limitations of this study is that our analysis of transcriptional changes due to abatacept could have been confounded by the specific immune activation that is due to RA. Like SARS, however, RA is a disease where macrophage IL6 production is central to the pathology (58). Of relevance, there is increasing evidence that abatacept could be an efficacious therapy to treat interstitial lung disease in RA (ILD) (59). ILD is a comorbidity that occurs in RA patients and is characterized by the inflammation and subsequent fibrotic development of the lung tissue. Treatment with abatacept was motivated by the finding that blocking T cell co-stimulation through CTLA4-Ig is effective in a mouse model of hypersensitivity pneumonitis (60), a disease characterized by a massive influx of activated T cells in the lungs and where alveolar macrophages express high levels of B7 (CD80/CD86) molecules. This evidence is in line with our hypothesis that the CD80/86 axis is a key factor for lung hyperinflammatory response. Additionally, while immunosuppressive agents are generally associated with an increase in the risk of infectious diseases in rheumatic patients (61), abatacept has shown to be the therapy that least increases this risk. There is substantial epidemiological evidence supporting that abatacept-treated patients do not increase the risk of serious infections (62)(63)(64).

Our finding that abatacept-treated patients have a lower percentage of COVID-19 symptoms, prompted us to provide additional evidence of the benefit of this therapeutic approach. After characterizing the transcriptional changes induced by the therapy in a cohort of RA patients, we have seen that it is highly antagonistic to that induced by COVID-19, both to several of the most well-known pathological processes associated with the disease, and to the biological processes that are differentially activated in samples of patients. Together, this evidence identifies blocking of the CD80/86 axis as a candidate therapy for COVID-19. To this regard, Belatacept, a CTLA4-Ig drug derived from abatacept that binds more avidly to CD80/86 and is used to prevent renal transplant rejection (65), could be a therapeutic alternative. However, there's yet no epidemiological evidence evaluating this drug in relation to COVID-19, with only sporadic reporting (66). While the development of vaccines and the questions like permanency of immunization are still being investigated, there is a need to find therapies that can lower the risk of progressing to severe stages of COVID-19 and reduce the number of fatalities during the next months or years. Additional clinical and experimental evidence gathered in the next months will be crucial to confirm if blocking of the CD80/86 axis is a useful therapeutic approach to prevent progression to severe stages of the disease.

**Acknowledgements**

The PACTABA project was funded Bristol-Myers Squibb. We thank all participants from the PACTABA study for their collaboration. AJ and SM are supported by the DoCTIS project funded by the European Union’s H2020 programme (grant #848028). This work was supported by funds from the Vall d’Hebron Hospital Research Institute and from IMIDomics. S.L. AJ, RMM and SM are also affiliated with IMIDomics.

**Author Contributions**

AJ directed the project, conceived, designed and analyzed data and wrote the manuscript; IB developed computational approaches, implemented bioinformatic analyses and wrote the manuscript; AG performed various computational analyses and wrote the manuscript; MLL and MLC contributed to patient recruitment, clinical data collection; SHMM performed statistical analyses; JLL contributed to data processing; INR performed sample sequencing experiments and revised the manuscript; RMM contributed to data generation and revised the manuscript; SM directed the RA project, conceived the project and wrote the manuscript.

**Declaration of Interests**

SM and RMM are co-founders of IMIDomics, a biotech company focused in bringing precision medicine to immune-mediated inflammatory diseases.

## Data and code availability

The whole code used to analyze the PBMC RNA-Seq data from COVID-19 infected patients, blood RNA-Seq data from RA patients treated with abatacept and the BALF single-cell RNA-Seq dataset is freely available at https://github.com/Rheumatology-Research-Group/COVID-19_Abatacept. The dataset generated during this study will be deposited in the Gene Expression Omnibus (GEO) repository.

## Figure Legends

**Figure 1. CTLA4-Ig (abatacept) dampens T cell co-stimulation and excessive cytokine production in COVID-19.** Schematics depicting the model for abatacept-mediated inhibition of excess cytokine production. Once lungs become infected with SARS-CoV2 virus, an immune response is mounted. Antigen presenting cells (APCs) in the lung present antigen to activated CD4+ T cells thereby stimulating their production of proinflammatory cytokines (e.g. TNF, IL6). Resident alveolar macrophages and, mostly, large numbers of infiltrating monocyte-derived macrophages respond to the T cell stimulation by producing large quantities of interleukins, mainly IL6. Large numbers of activated macrophages present in the lung can lead to systemic secretion of IL6, forming the cytokine storm. Treatment with CTLA4-Ig recombinant fusion protein could dampen APC activation of T cells and, therefore, reduce the level of hyperinflammatory response observed in severe COVID-16 patients.

**Figure 2. Longitudinal transcriptional changes induced by CTLA4-Ig therapy in COVID-19 associated processes. A.** Changes in biological processes associated with immune viral response. **B.** Changes in biological processes associated with hyperinflammation and severity in COVID-19. The first column on the left indicates the direction of change of the biological process by COVID19 as described in different clinical and experimental studies (blue: up-regulation, red: down-regulation). As a reference, we include the transcriptional changes observed for the same set of processes in the COVID-19 PBMC samples (middle column). Finally, the third column shows the change induced by treatment with abatacept in RA patients. Up-regulation (blue) and down-regulation (red) is depicted according to effect size (diameter of circle). Statistical significance is reflected as color intensity and p-value levels according to *<0.05, **<0.005, ***<0.0005, ****<0.00005 and *****<0.000005.

**Figure 3. Gene-level differential expression of the COVID-19 related biological processes induced by abatacept. A.** Immune sensing and response to viral infection. **B.** Immune processes associated with COVID-19 severity stages. Volcano-plots showing the statistical significance (-log10(p-value), y-axis) against the effect size (log fold change, x-axis) for all genes. In color are highlighted the genes from the specific biological processes, red for genes down-regulated by abatacept, and blue for genes up-regulated by abatacept.

**Figure 4. Venn plot of the biological processes regulated by COVID-19 and by abatacept.** Numbers represent the biological processes that are specific and those that are commonly changed by both exposures. From the latter, 47 processes modified by COVID-19 are antagonistically modified by abatacept, compared to only 2 processes that are agonistically induced by co-stimulation inhibition.

**Figure 5. Most significant COVID-19 transcriptional processes antagonized by abatacept.** The 20 top biological processes more significantly down or up-regulated by COVID-19 at the transcriptional level and also more significantly antagonized by abatacept are shown in this diagram. Up-regulation (blue) and down-regulation (red) is depicted according to effect size (diameter of circle). Statistical significance is reflected as color intensity and p-value levels according to *<0.05, **<0.005, ***<0.0005, ****<0.00005 and *****<0.000005.

**Figure 6. UMAP of the BALF samples from COVID-19 and healthy control individuals.** Total cell BALF samples from 6 severe and 3 mild COVID-19 infected individuals and 3 CD45+ sorted BALF samples from 3 healthy controls were analyzed for cell cluster identification. Each number depicts each of the 34 cell clusters identified using the SNN approach implemented in Seurat v3.

**Figure 7. Expression of cell type specific markers in BALF samples from COVID-19 and healthy controls.** 2-Dimensional visualization of gene expression based on the identified cell clusters (Figure 6). **A.** Previously described markers FCN1+ and FABP4+ defining monocyte-derived macrophages and resident alveolar macrophages, respectively. The two cluster aggregates that entirely express FABP4 correspond of cells from healthy controls. **B.** Expression of co-stimulating protein genes CD80 and CD86, which are the target of abatacept. While CD86 has a more ubiquitous expression (including both FCN1+ and FABP4+ macrophages), CD80 is specifically expressed in FCN1+ monocyte-derived macrophages. **C.** IL6 expression is circumscribed to the activated macrophage cluster. **D.** Selected T cell markers representing activated T cells (CTLA4, IL2RA), CD8+ lymphocytes (CD8A) and Tregs (FOXP3 and IL2RA). CTLA4 expression is enriched in the CD4+ T cell cluster 5.

**Figure 8. CD80/86 axis cell type abundance and correlation.** Scatter plots showing the cell type abundance correlation between the key elements of the CD80/86 axis targeted by CTLA4-Ig therapy. All pairwise correlation analyses were found to be positive and significant ($r^2 > 0.7$, $P<0.005$). **A** CD86+ cells vs active CD4+ T cells. **B** CD80+ cells vs active CD4+ T cells. **C** active CD4+ T cells vs IL6+ cells. **D** CD86+ cells vs IL6+ cells. **E** CD80+ cells vs IL6+ cells.

## Supplementary Figure Legends

**Supplementary Figure 1. Evidence of neutrophil presence in the COVID-19 PBMC dataset.** Using the cell type deconvolution approach in ABIS, we identified a potential presence of neutrophils in the PBMC samples. In order to correct for this potential confounder, we used the estimated neutrophil percentage as covariate in the differential expression analysis. N1-3: control samples; P1-P3: patient samples.

**Supplementary Figure 2. UMAP depicting the individual origin of the cell clusters.** Cell coloring is based on the 12 individuals simultaneously analyzed in the BALF scRNA-Seq study. Samples C51, C52 and C100 correspond to the three controls, which clearly show distinct clusters of FABP4+ alveolar macrophage cells. The remaining samples are from COVID-19 patients with 3 samples from mild patients (C141, C142 and C144) and 6 samples from patients with severe disease (C143, C145, C146, C148, C149, C152).

**Supplementary Figure 3. Tissue-level gene expression distribution for CD80, CD86 and CTLA4 genes from GTex database.** The distribution of gene expression levels for the three key genes from 53 human tissues from nearly 1,000 individuals are shown (GTex database, version 8). All three key genes from the CD80/86 axis are highly expressed in the lung, being the first tissue for *CD80*, the second for *CD86* and the third for *CTLA4* mRNAs.

**Fig. 1**

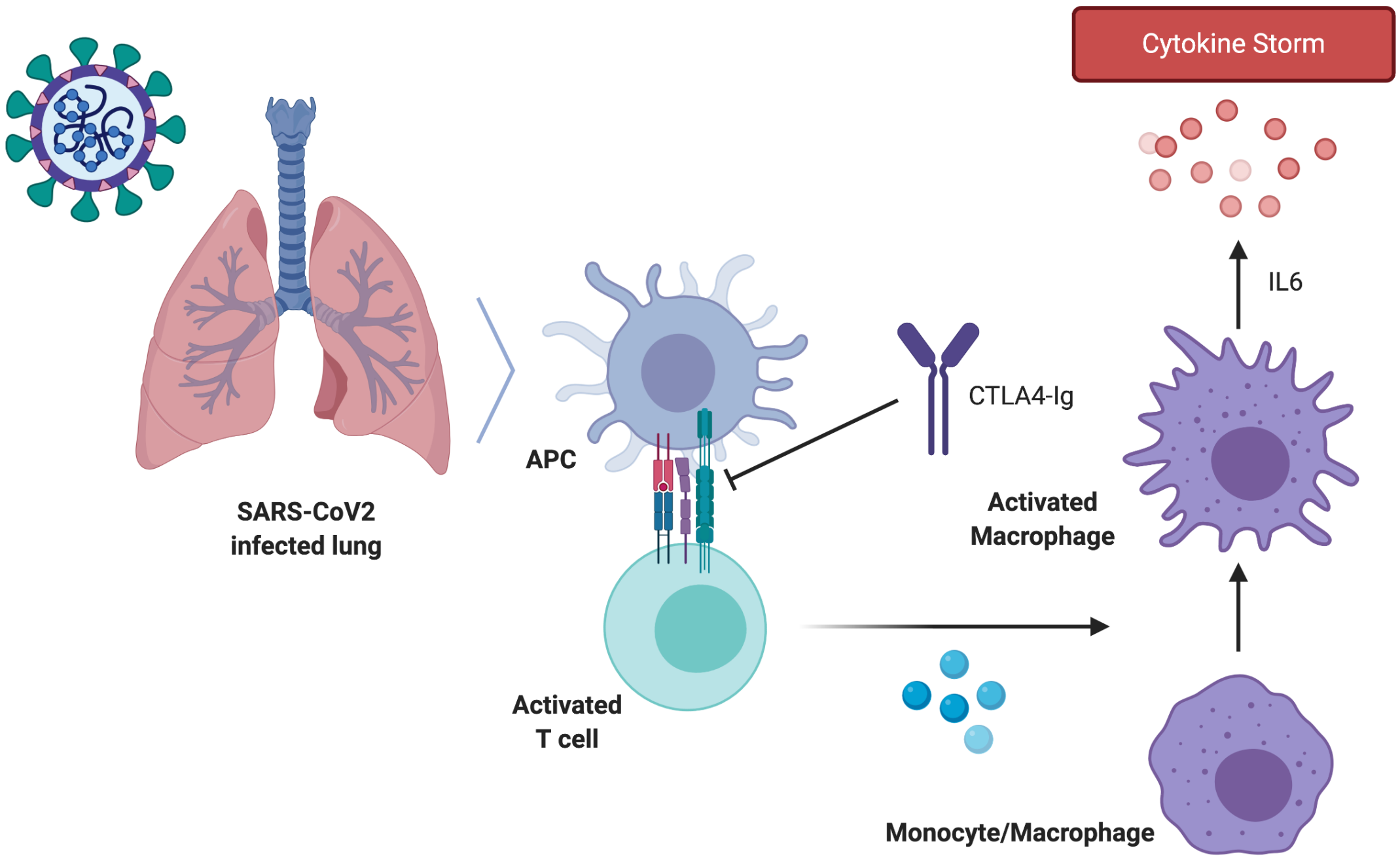

Cytokine Storm
IL6
SARS-CoV2
infected lung
APC
CTLA4-Ig
Activated
Macrophage
Activated
T cell
Monocyte/Macrophage

**Fig. 2 A**

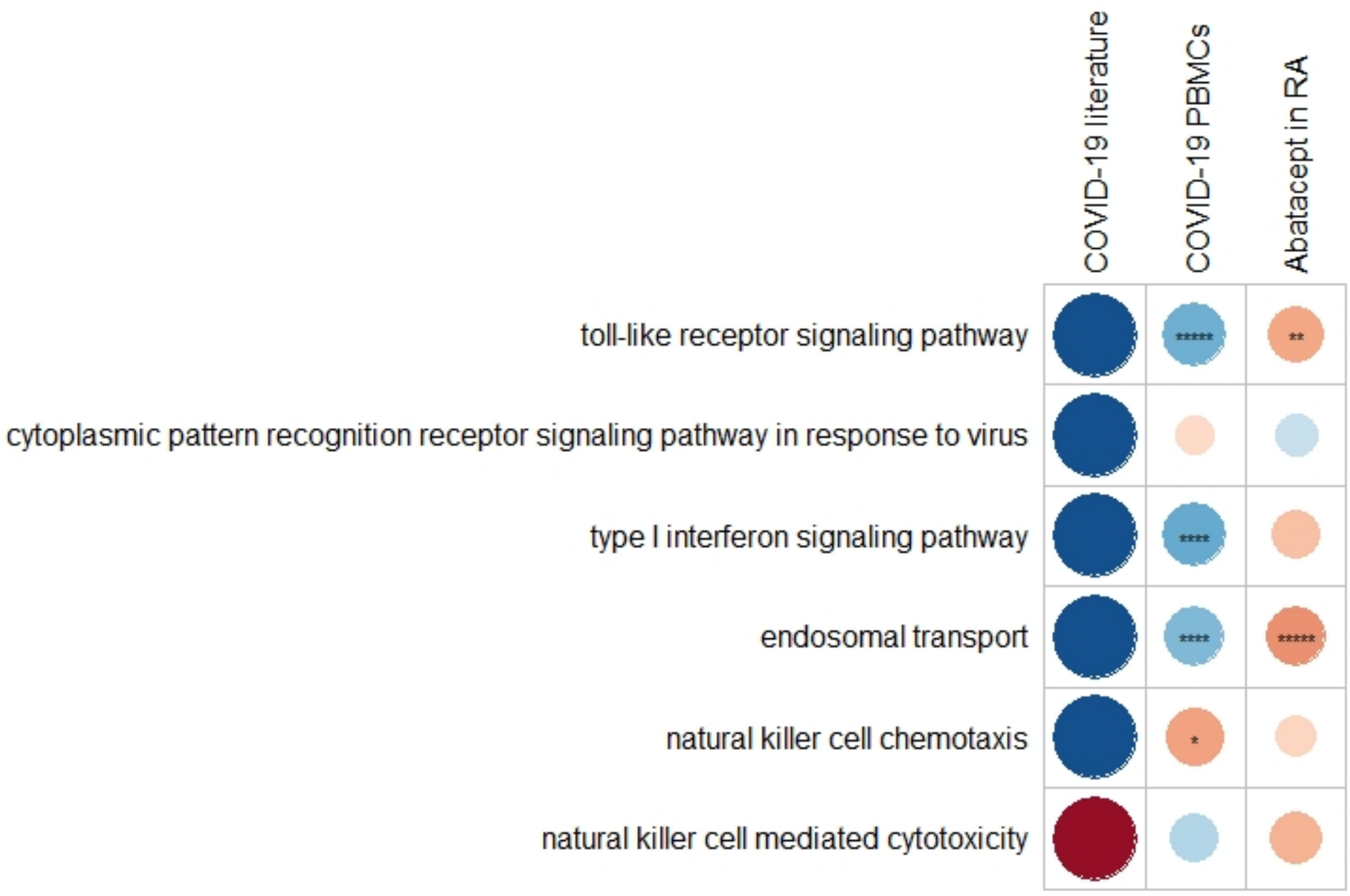

COVID-19 literature
COVID-19 PBMCs
Abatacept in RA
toll-like receptor signaling pathway
cytoplasmic pattern recognition receptor signaling pathway in response to virus
type I interferon signaling pathway
endosomal transport
natural killer cell chemotaxis
natural killer cell mediated cytotoxicity

**Fig. 2 B**

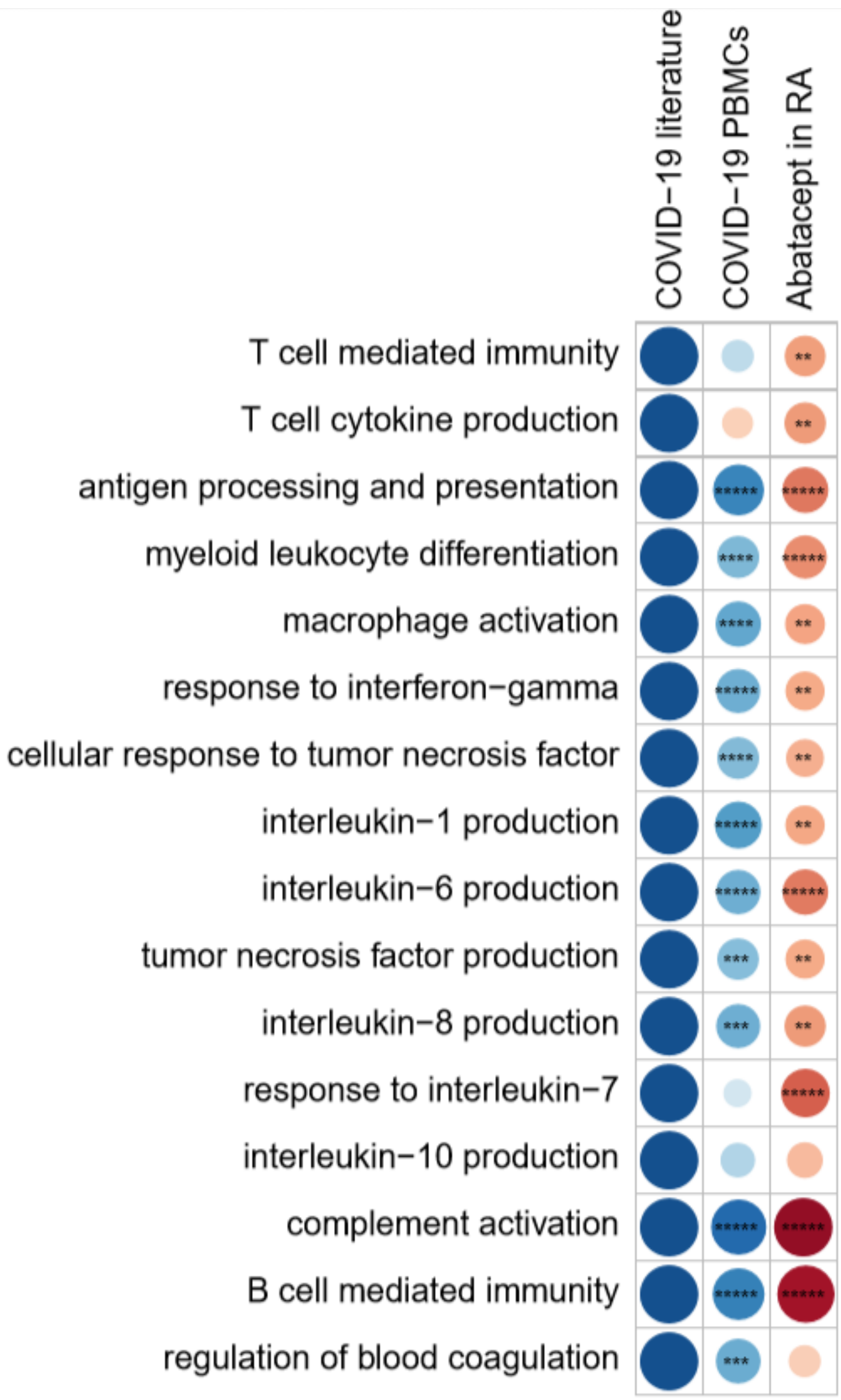

COVID−19 literature
COVID−19 PBMCs
Abatacept in RA
T cell mediated immunity
T cell cytokine production
antigen processing and presentation
myeloid leukocyte differentiation
macrophage activation
response to interferon−gamma
cellular response to tumor necrosis factor
interleukin−1 production
interleukin−6 production
tumor necrosis factor production
interleukin−8 production
response to interleukin−7
interleukin−10 production
complement activation
B cell mediated immunity
regulation of blood coagulation

**Fig. 3. A**

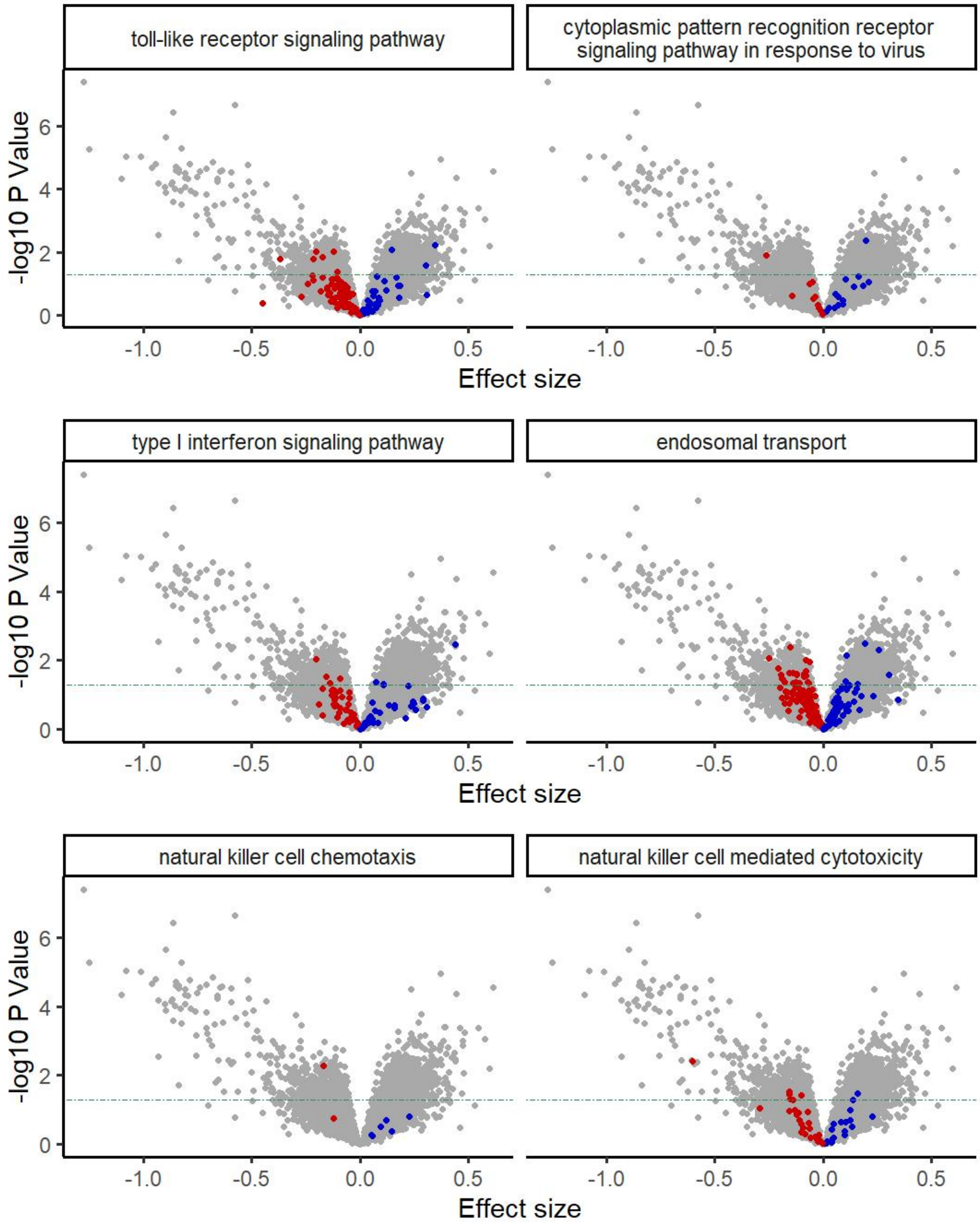

toll-like receptor signaling pathway
cytoplasmic pattern recognition receptor signaling pathway in response to virus
type I interferon signaling pathway
endosomal transport
natural killer cell chemotaxis
natural killer cell mediated cytotoxicity
-log10 P Value
0
2
4
6
-1.0
-0.5
0.0
0.5
Effect size

**Fig. 3. B**

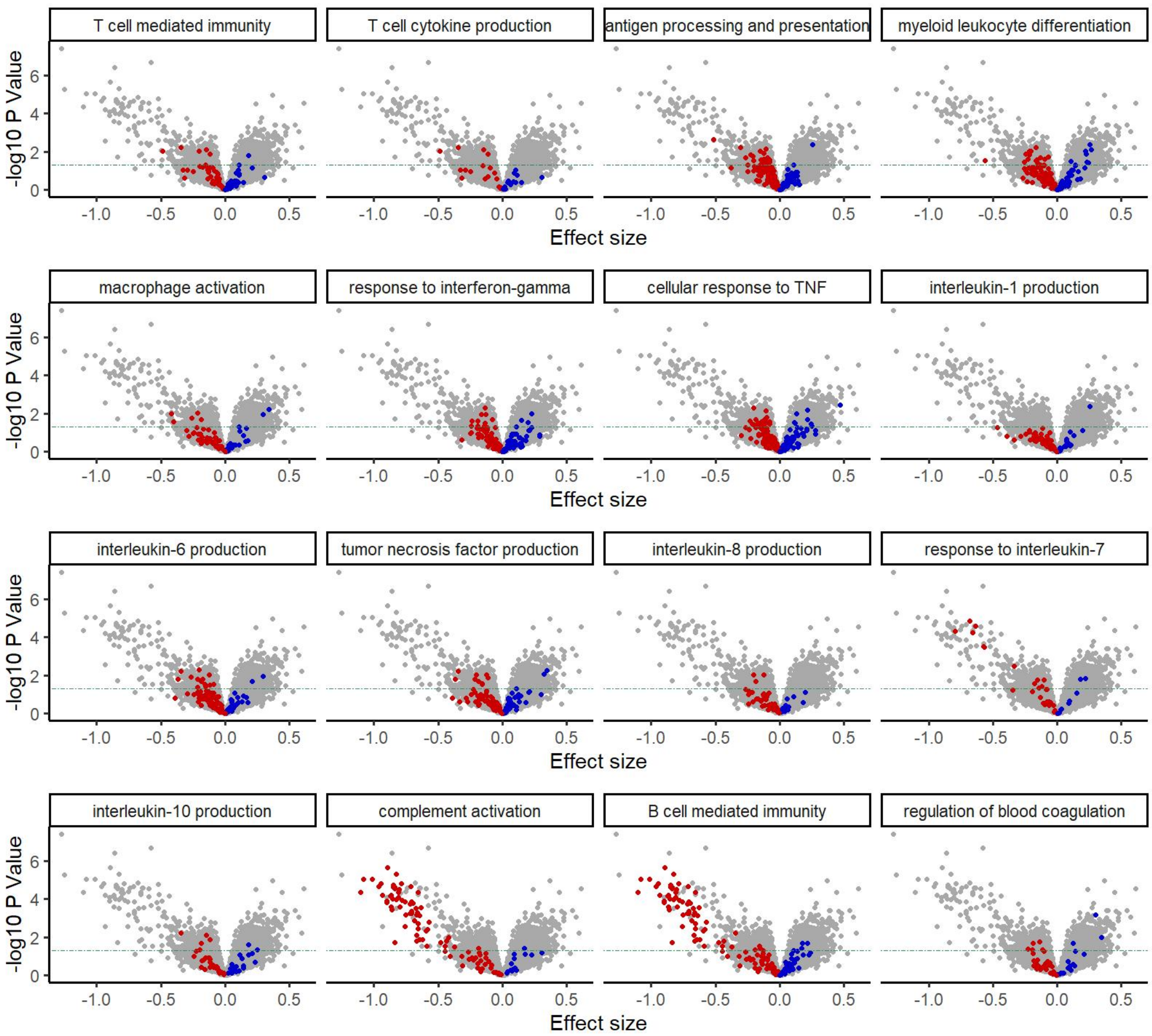

T cell mediated immunity
T cell cytokine production
antigen processing and presentation
myeloid leukocyte differentiation
macrophage activation
response to interferon-gamma
cellular response to TNF
interleukin-1 production
interleukin-6 production
tumor necrosis factor production
interleukin-8 production
response to interleukin-7
interleukin-10 production
complement activation
B cell mediated immunity
regulation of blood coagulation
-log10 P Value
0
2
4
6
Effect size
-1.0
-0.5
0.0
0.5

**Fig. 4**

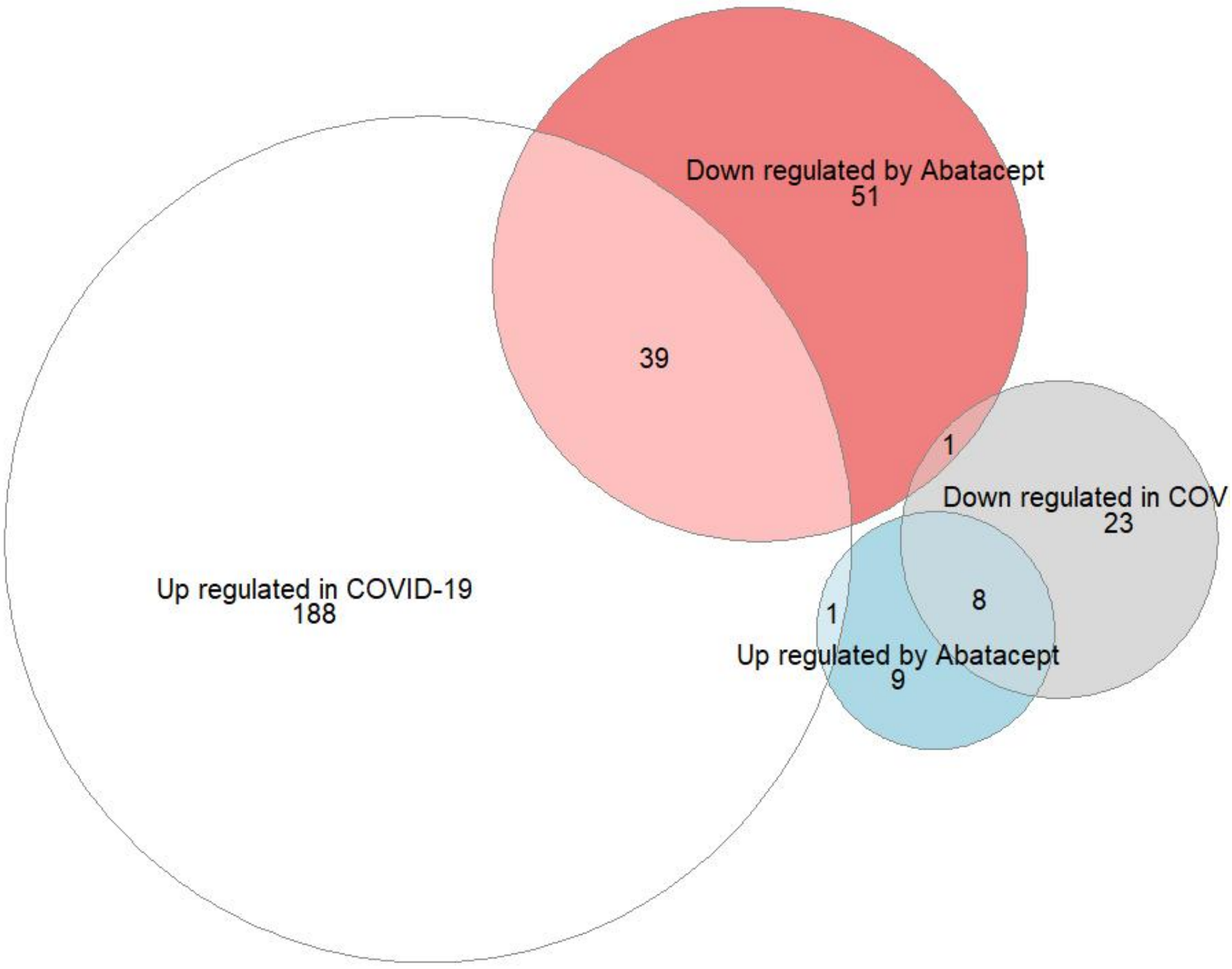
Down regulated by Abatacept
51
39
1
Down regulated in COV
23
Up regulated in COVID-19
188
1
8
Up regulated by Abatacept
9

Fig. 5

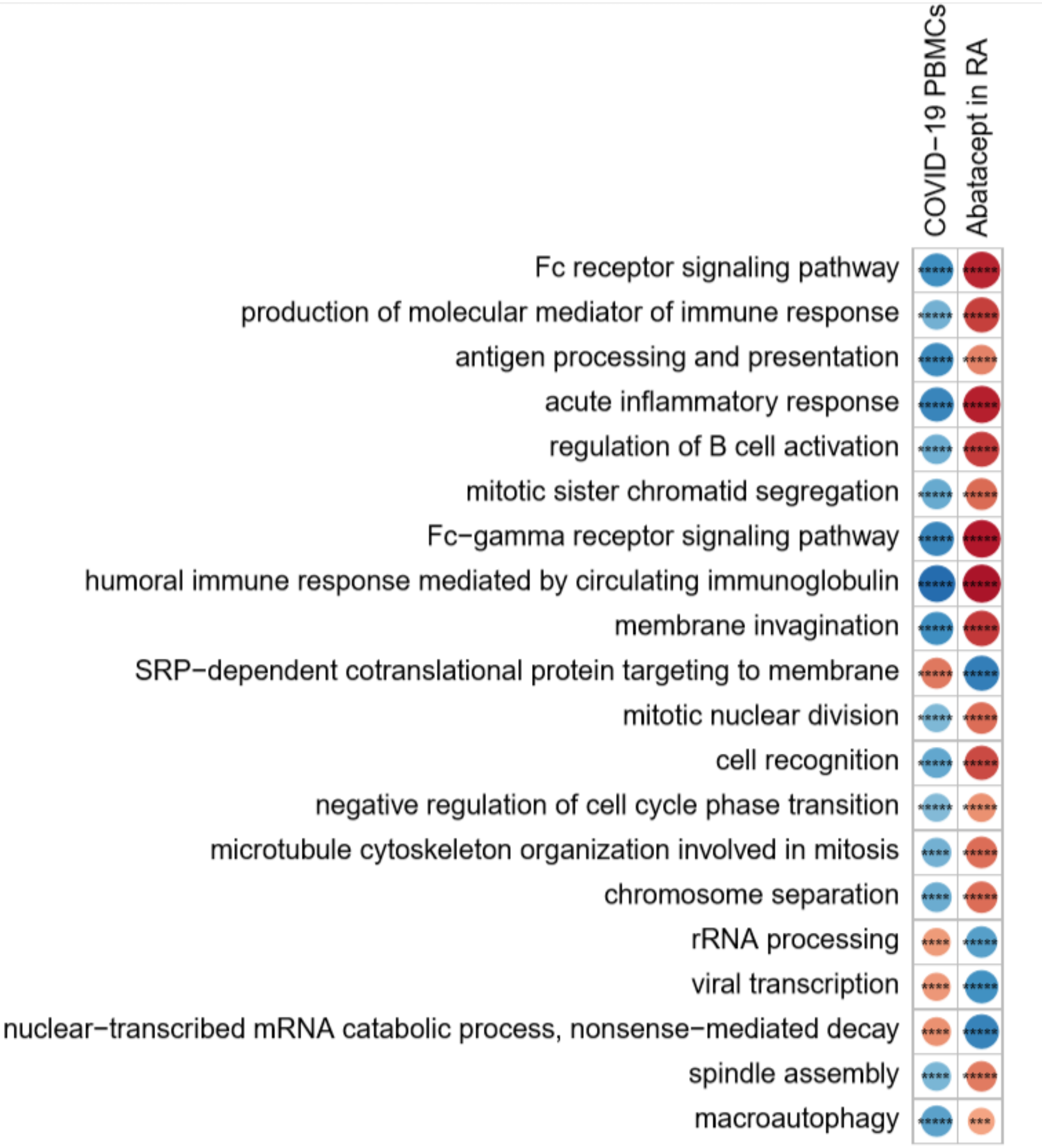
COVID−19 PBMCs
Abatacept in RA
Fc receptor signaling pathway
production of molecular mediator of immune response
antigen processing and presentation
acute inflammatory response
regulation of B cell activation
mitotic sister chromatid segregation
Fc−gamma receptor signaling pathway
humoral immune response mediated by circulating immunoglobulin
membrane invagination
SRP−dependent cotranslational protein targeting to membrane
mitotic nuclear division
cell recognition
negative regulation of cell cycle phase transition
microtubule cytoskeleton organization involved in mitosis
chromosome separation
rRNA processing
viral transcription
nuclear−transcribed mRNA catabolic process, nonsense−mediated decay
spindle assembly
macroautophagy

**Fig. 6**

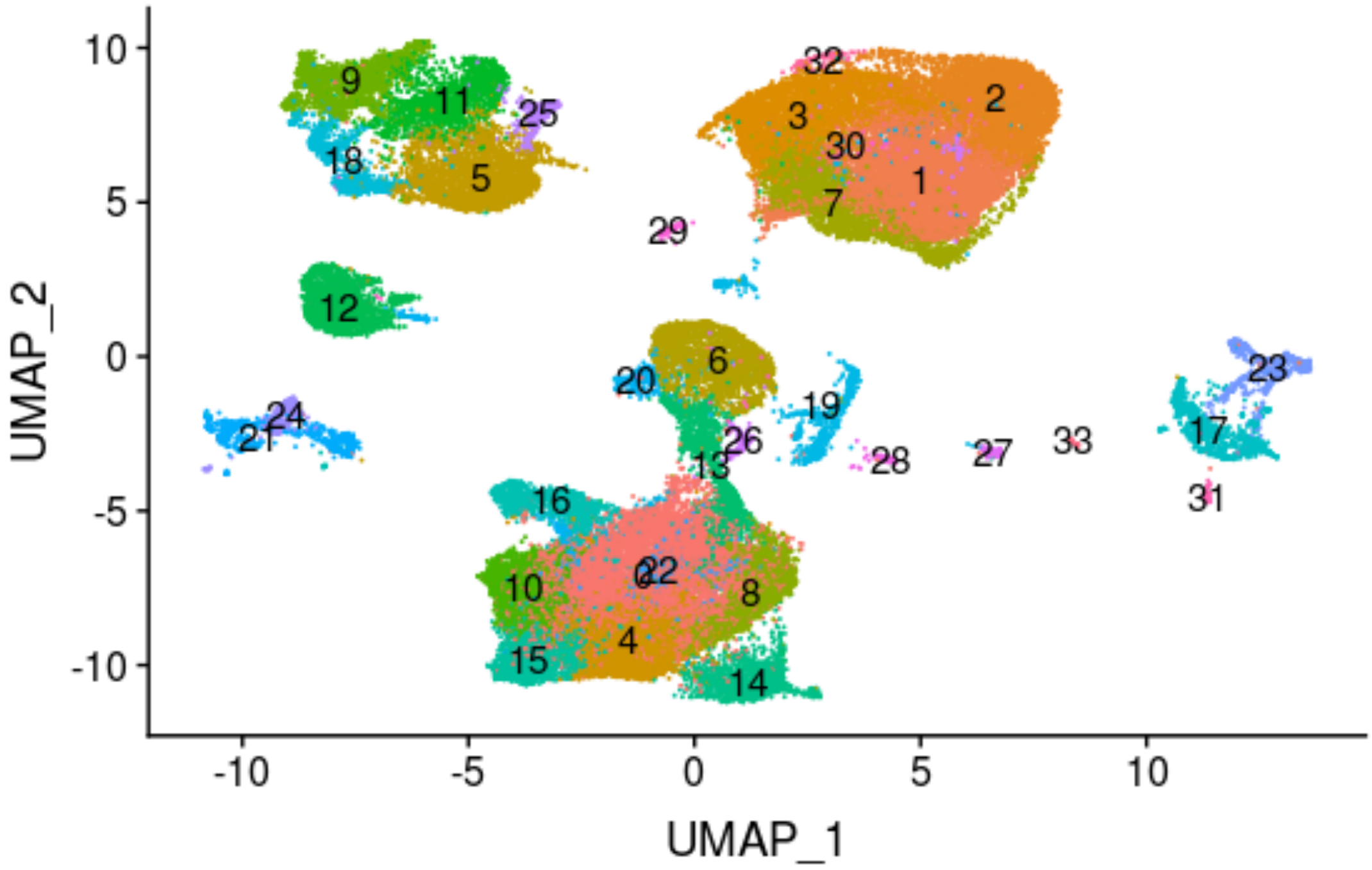
UMAP_2
UMAP_1
10
5
0
-5
-10
-10
-5
0
5
10
9
11
25
18
5
12
24
21
32
3
2
30
1
7
29
6
20
19
26
13
28
27
33
23
17
31
16
22
10
8
4
15
14

**Fig. 7 A**

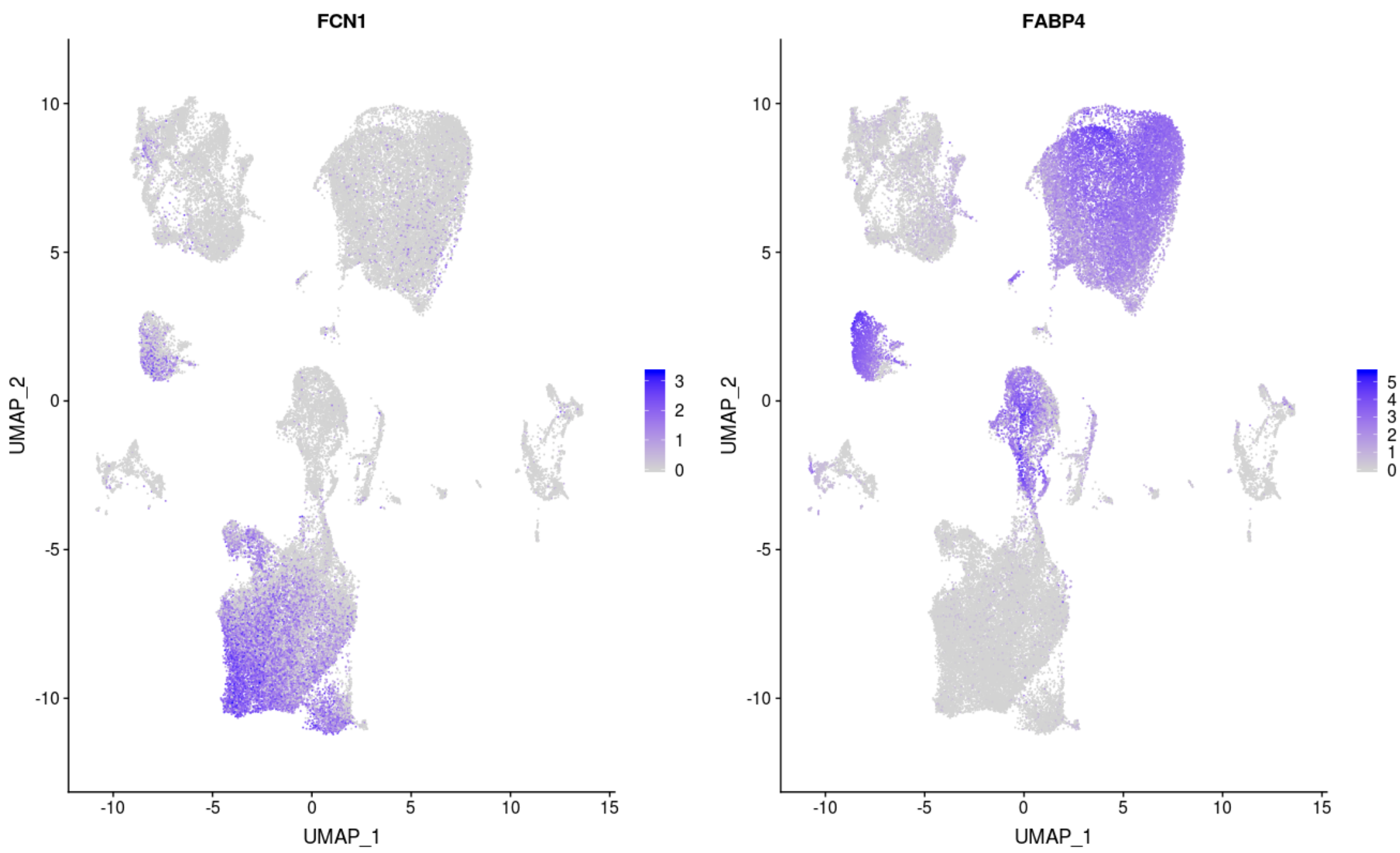
FCN1
FABP4
UMAP_2
UMAP_1
10
5
0
-5
-10
-10
-5
0
5
10
15
3
2
1
0
5
4
3
2
1
0

**Fig. 7 B**

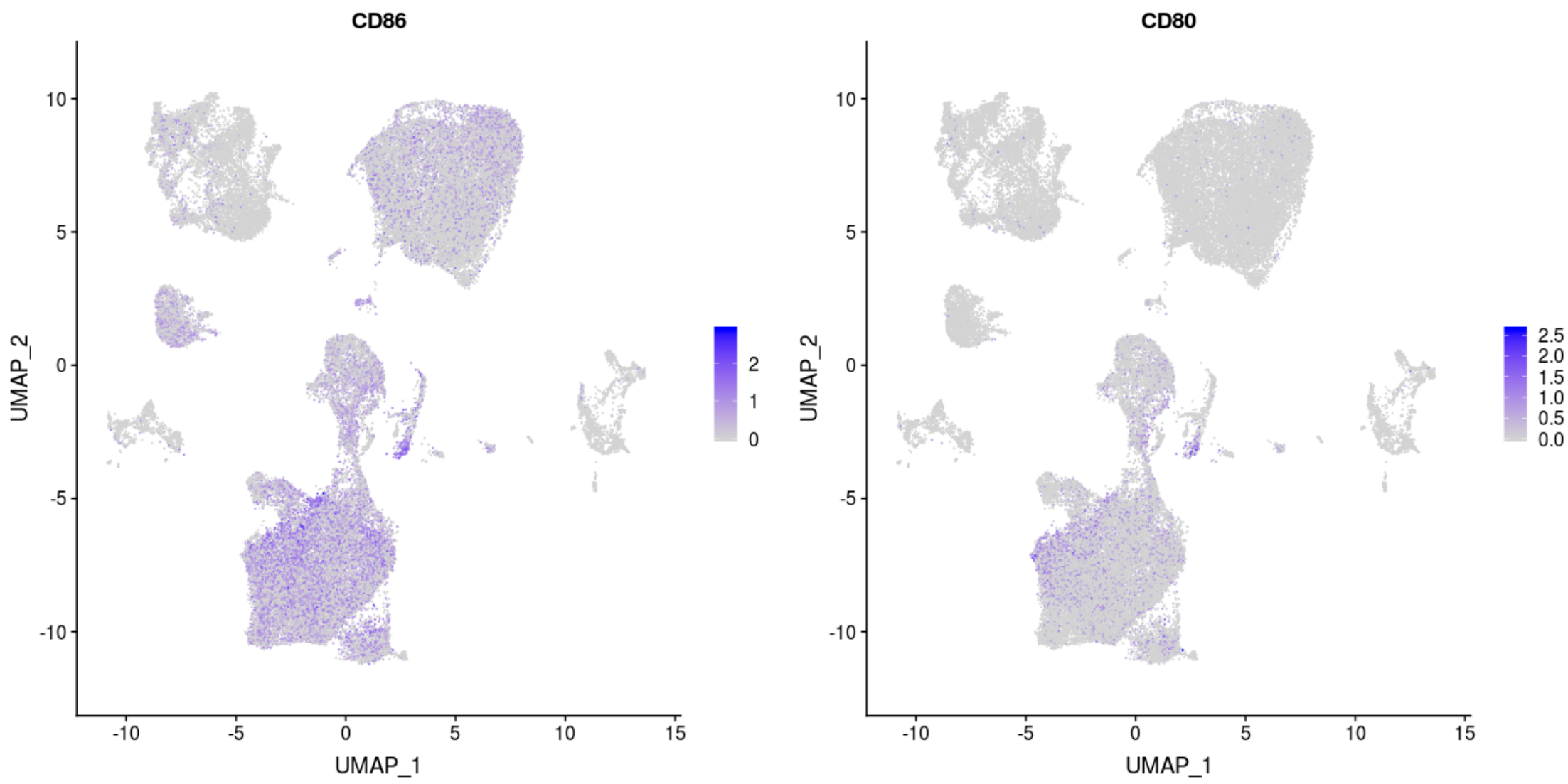
CD86
CD80
UMAP_1
UMAP_2
-10
-5
0
5
10
15
2.5
2.0
1.5
1.0
0.5
0.0

**Fig. 7 C**

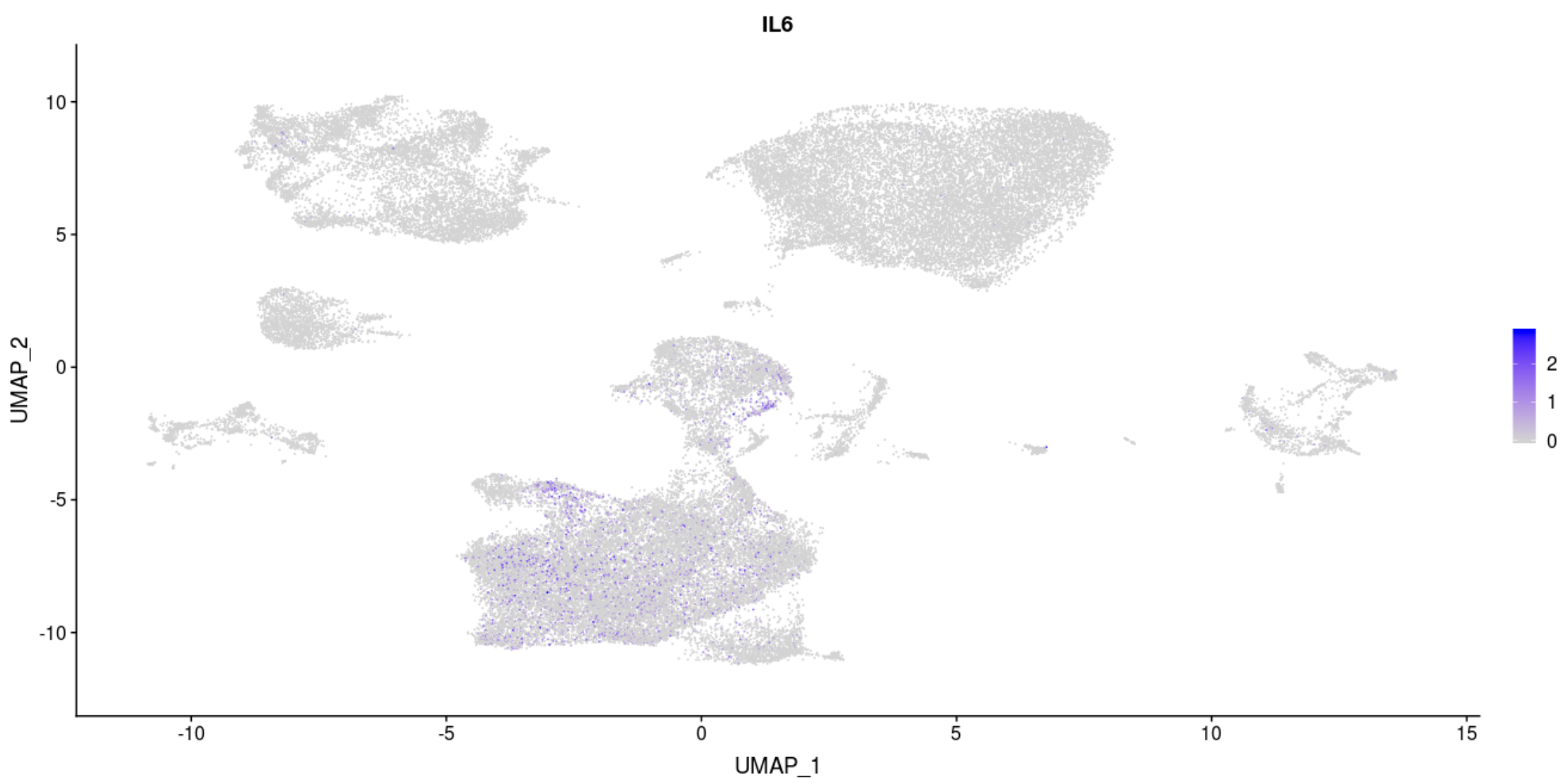
IL6
UMAP_1
UMAP_2
-10
-5
0
5
10
15
2
1
0

**Fig. 7 D**

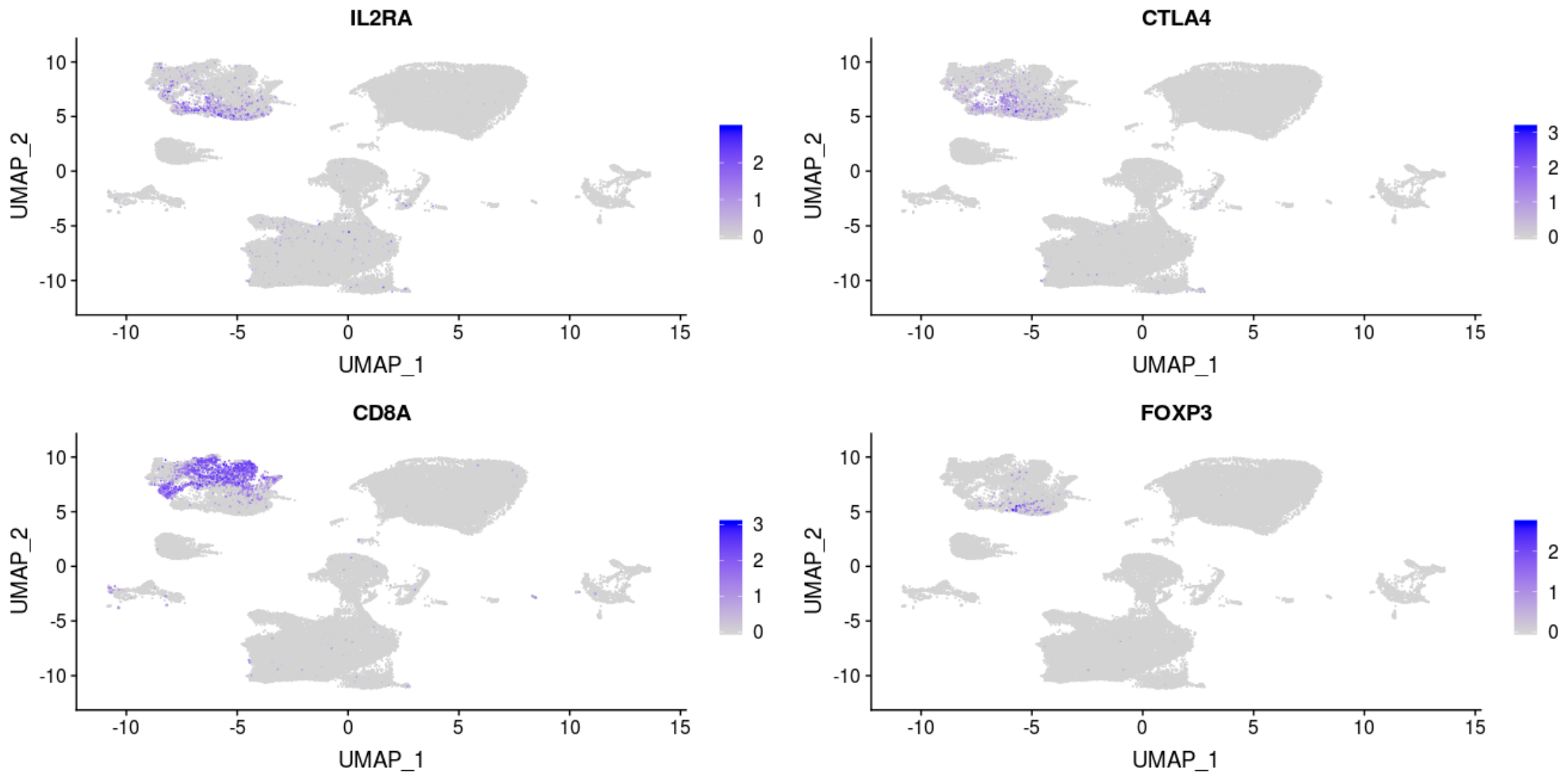
IL2RA
CTLA4
CD8A
FOXP3
UMAP_1
UMAP_2

**Fig. 8 A**

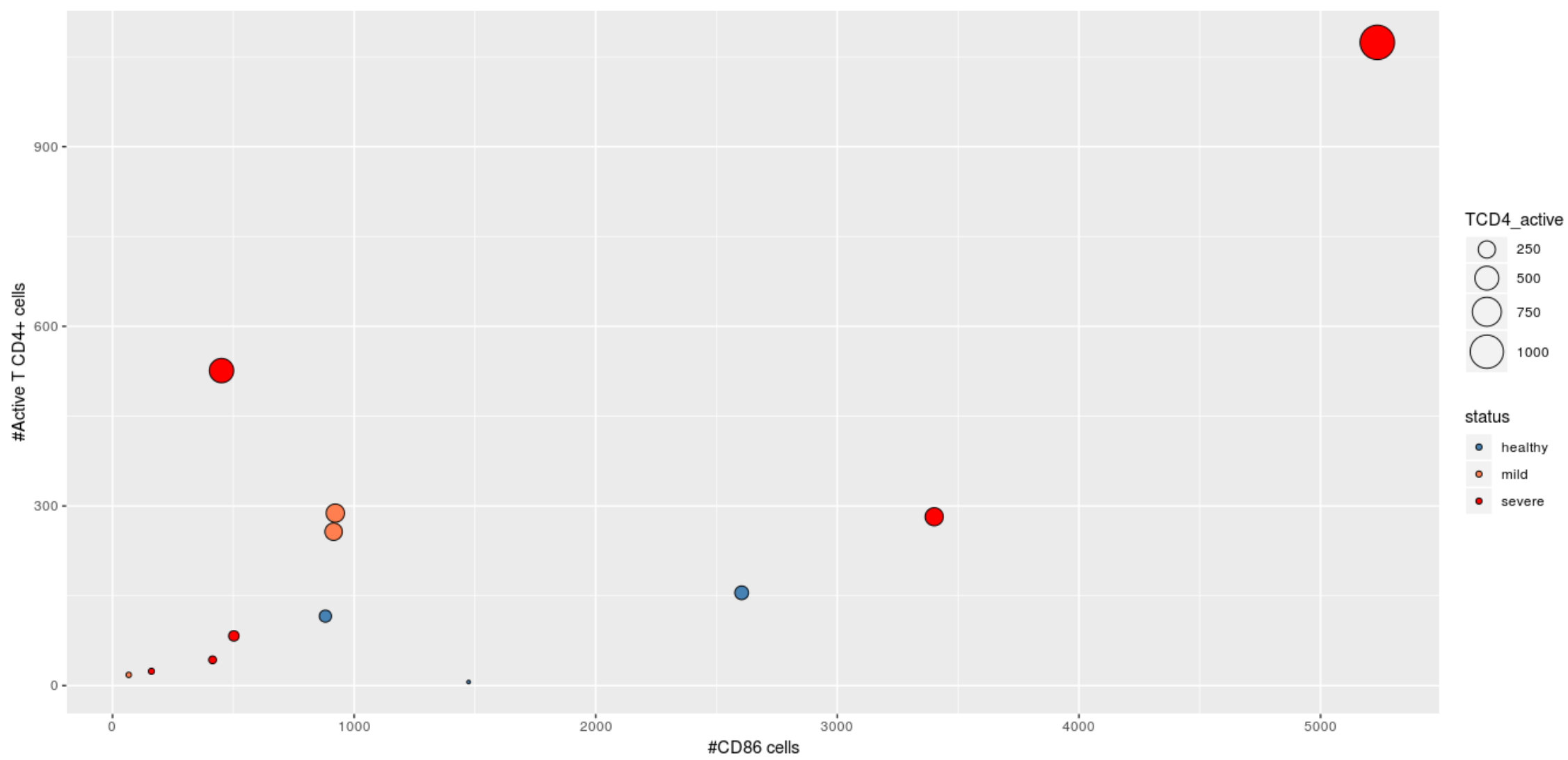
#Active T CD4+ cells
#CD86 cells
TCD4_active
250
500
750
1000
status
healthy
mild
severe

Fig. 8 B

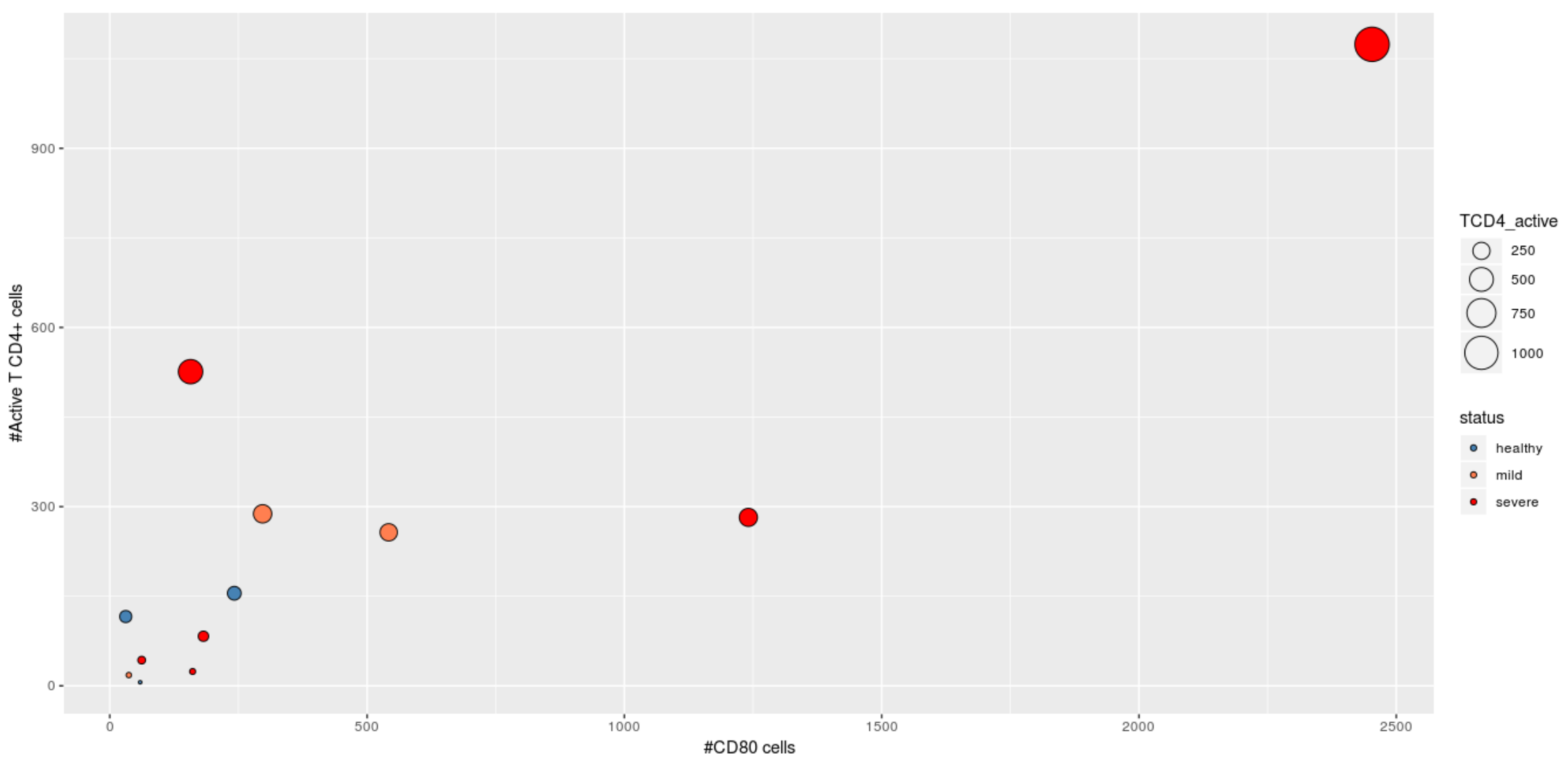

#Active T CD4+ cells
900
600
300
0
0
500
1000
1500
2000
2500
#CD80 cells
TCD4_active
250
500
750
1000
status
healthy
mild
severe

Fig. 8 C

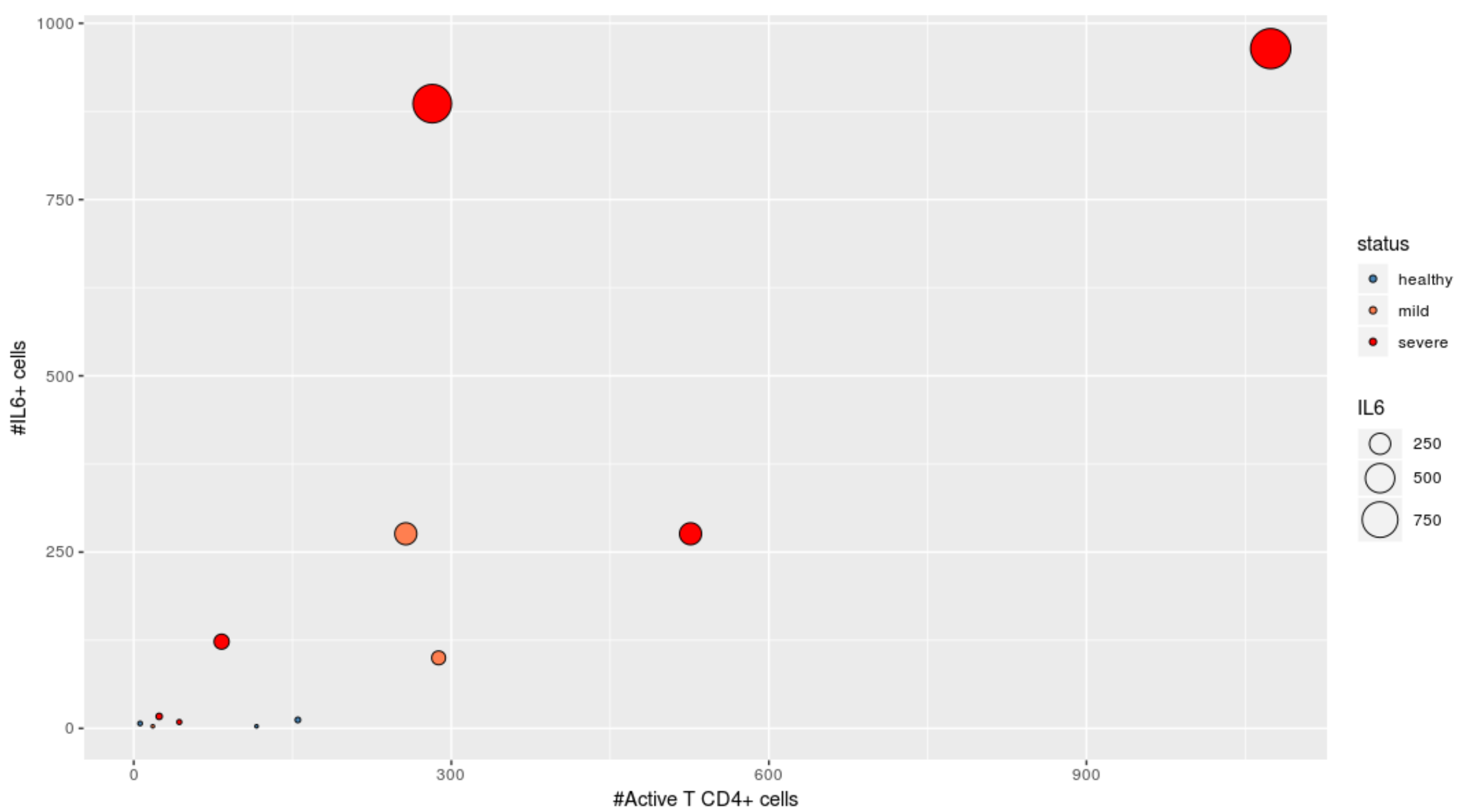

#IL6+ cells
1000
750
500
250
0
0
300
600
900
#Active T CD4+ cells
status
healthy
mild
severe
IL6
250
500
750

**Fig. 8 D**

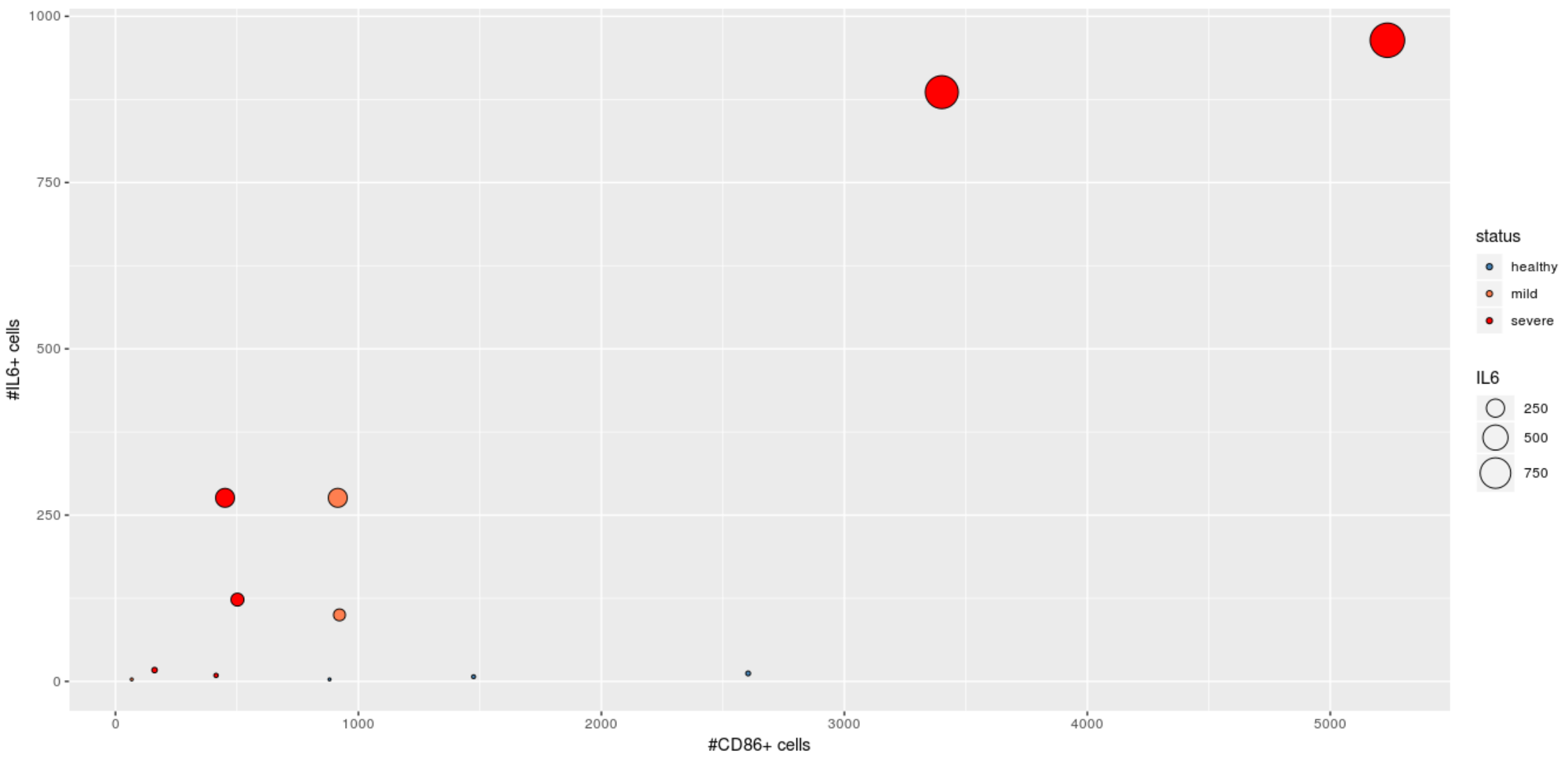
1000
750
500
250
0
#IL6+ cells
0
1000
2000
3000
4000
5000
#CD86+ cells
status
healthy
mild
severe
IL6
250
500
750

**Fig. 8 E**

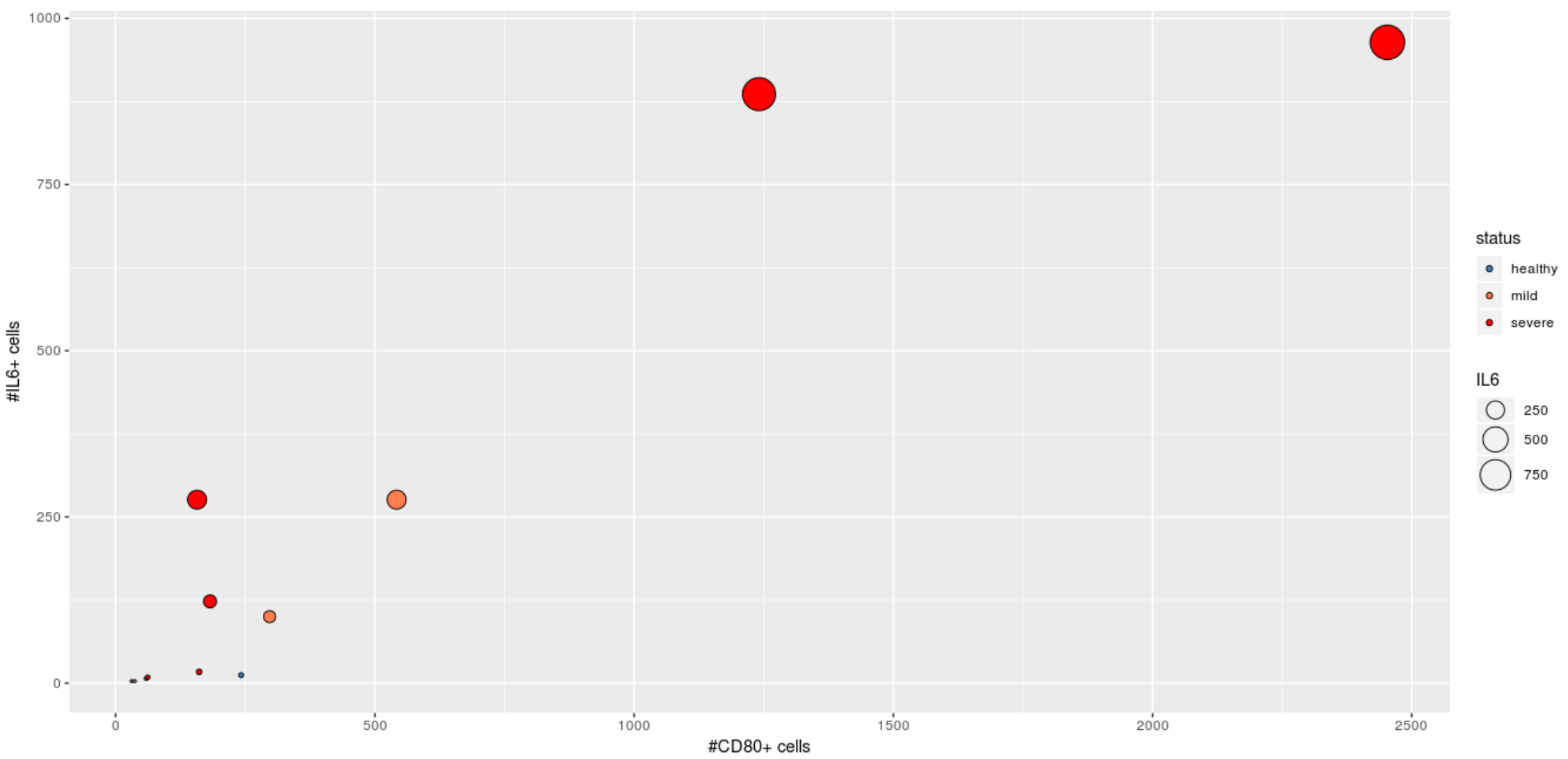
1000
750
500
250
0
#IL6+ cells
0
500
1000
1500
2000
2500
#CD80+ cells
status
healthy
mild
severe
IL6
250
500
750

**Table 1. List of biological processes associated with COVID-19 pathology**

| Biological process | GO term | Implication in SARS-CoV-2 | References |
|---|---|---|---|
| Viral entry into cell | endosomal transport | Mechanism of viral entry into cell. | (33)(67) |
| Virus sensing | toll−like receptor signaling pathway | Dificultates cell infection. Evaded by CoVs. | (33) |
| Virus sensing | cytoplasmic pattern recognition receptor signaling pathway in response to virus | Dificultates cell infection. Evaded by CoVs. | (33) |
| Virus sensing | type I interferon signaling pathway | Unclear. Impaired in severe. | (68)(5)(69) |
| Natural killer mediated immunity | natural killer cell chemotaxis | Over-activation | (17)(15) |
| Natural killer mediated immunity | natural killer cell mediated cytotoxicity | Down-regulated | (70)(71) |
| Blood coagulation | regulation of blood coagulation | Up regulated in severe. | (72)(73) |
| **Biological process** | **GO term** | **Implication in SARS-CoV-2** | **References** |
| T cells | T cell mediated immunity | Blood lymphopenia, altered function, activated and exhausted in severe. Increased in lung. | (8)(74)(74)(70)(75)(76)(77)(78)(79)(80)(17)(81) |
| T cells | T cell cytokine production | Over activation. Up in severe. | (8)(74)(80)(17)(81) |
| T cell interaction with myeloid cells | antigen processing and presentation | Up in severe | (44) |
| T cell interaction with myeloid cells | response to interferon−gamma | Over activation | (48)(8)(82) |
| T cell interaction with myeloid cells | cellular response to tumor necrosis factor | Over activation | (8) |
| Myeloid cell activation | myeloid leukocyte differentiation | Over activation. Up in severe. | (8)(74)(83)(76)(84)(17) |
| Myeloid cell activation | macrophage activation | Over activation. Up in severe. | (8)(74)(83)(76)(17)(73) |
| Cytokine production | interleukin−1 production | Over activation | (48)(76)(81) |
| Cytokine production | interleukin−6 production | Over activation. Up in severe. | (49)(8)(75)(69)(83)(84)(77)(73)(82) |
| Cytokine production | tumor necrosis factor production | Over activation. Up in severe. | (48)(49)(8)(75)(69)(83)(22)(73) |
| Cytokine production | interleukin−8 production | Over activation. Up in severe. | (48)(49)(75)(73) |
| Cytokine production | response to interleukin−7 | Over activation. Up in severe. | (48)(8) |
| Cytokine production | interleukin−10 production | Over activation. Up in severe. | (48)(49)(75)(77)(82) |
| Complement pathway | complement activation | Over activation in lung. Up in severe. | (32) |
| Ig production by B cells | B cell mediated immunity | Over activation. Up in severe. | (33)(51)(52)(80) |

**Suppl. Fig. 1**

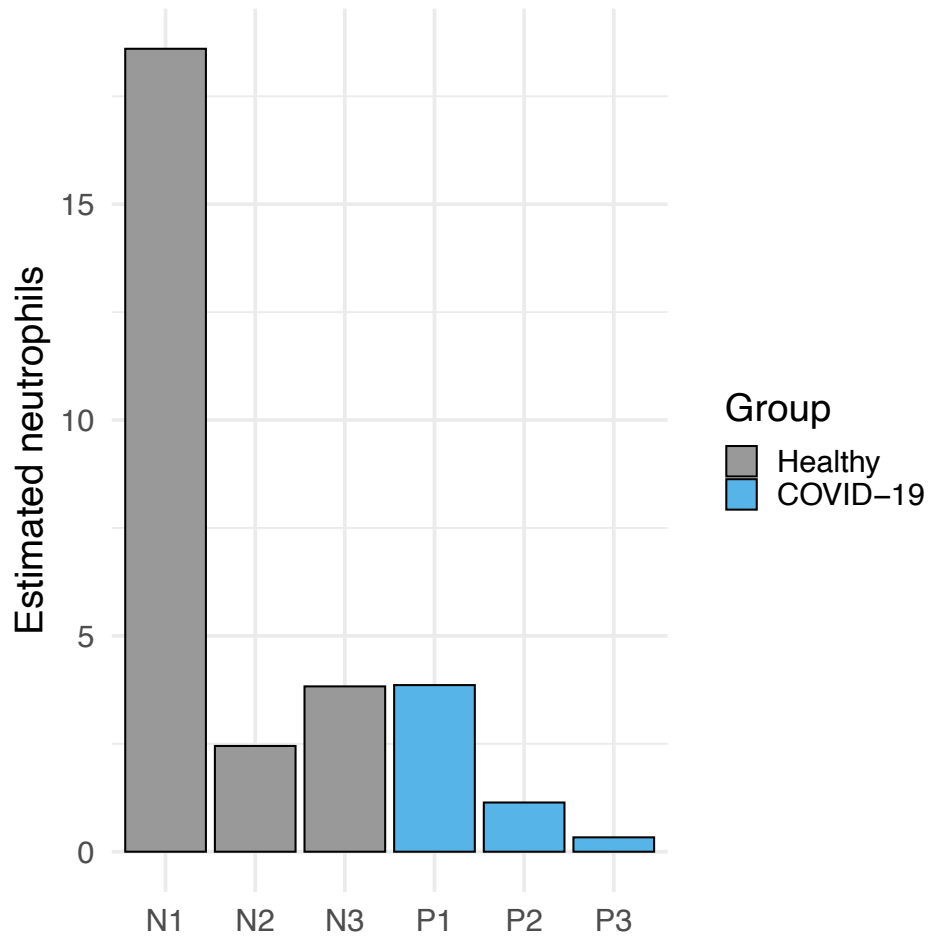

**Suppl. Fig. 2**

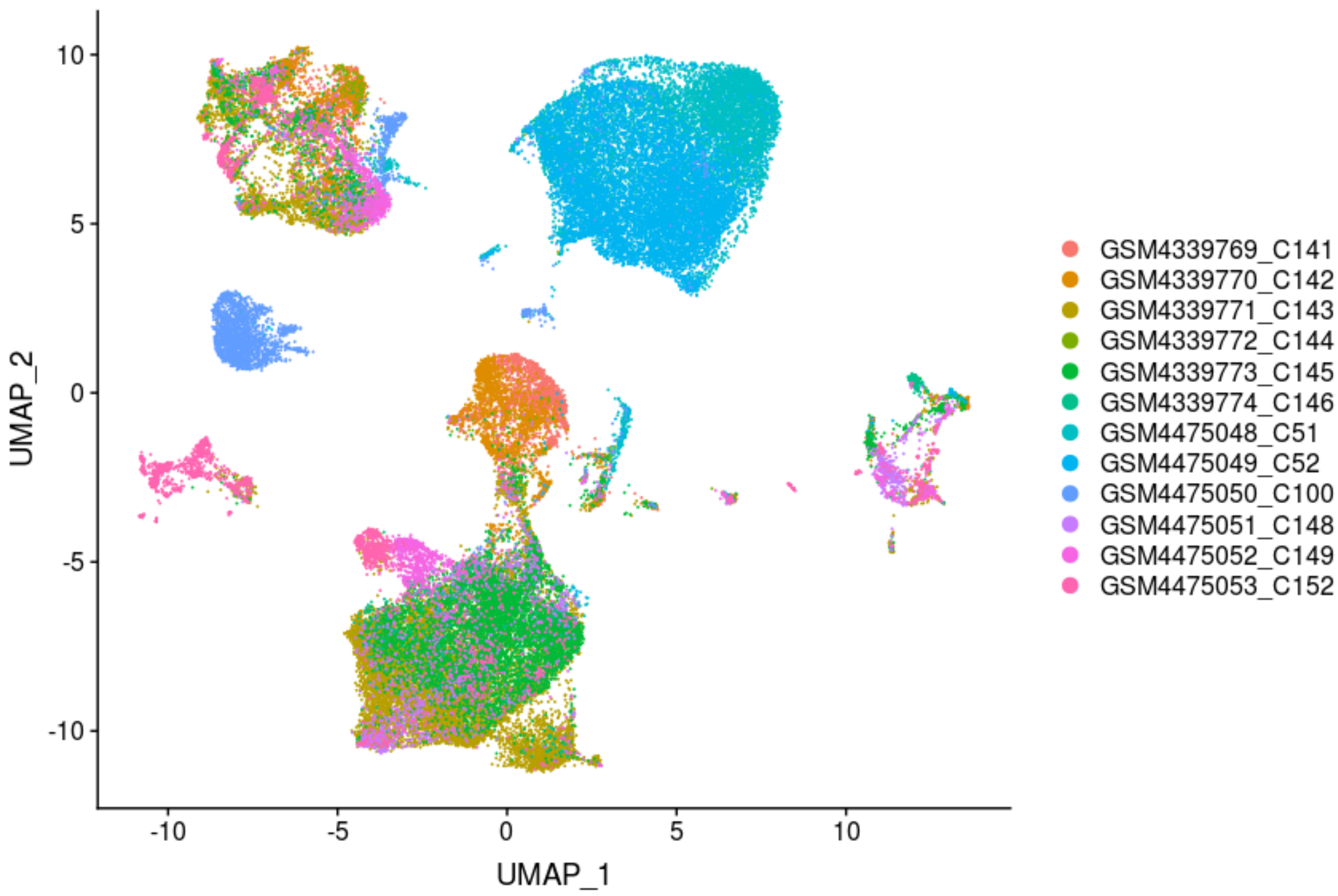

## Suppl. Fig. 3

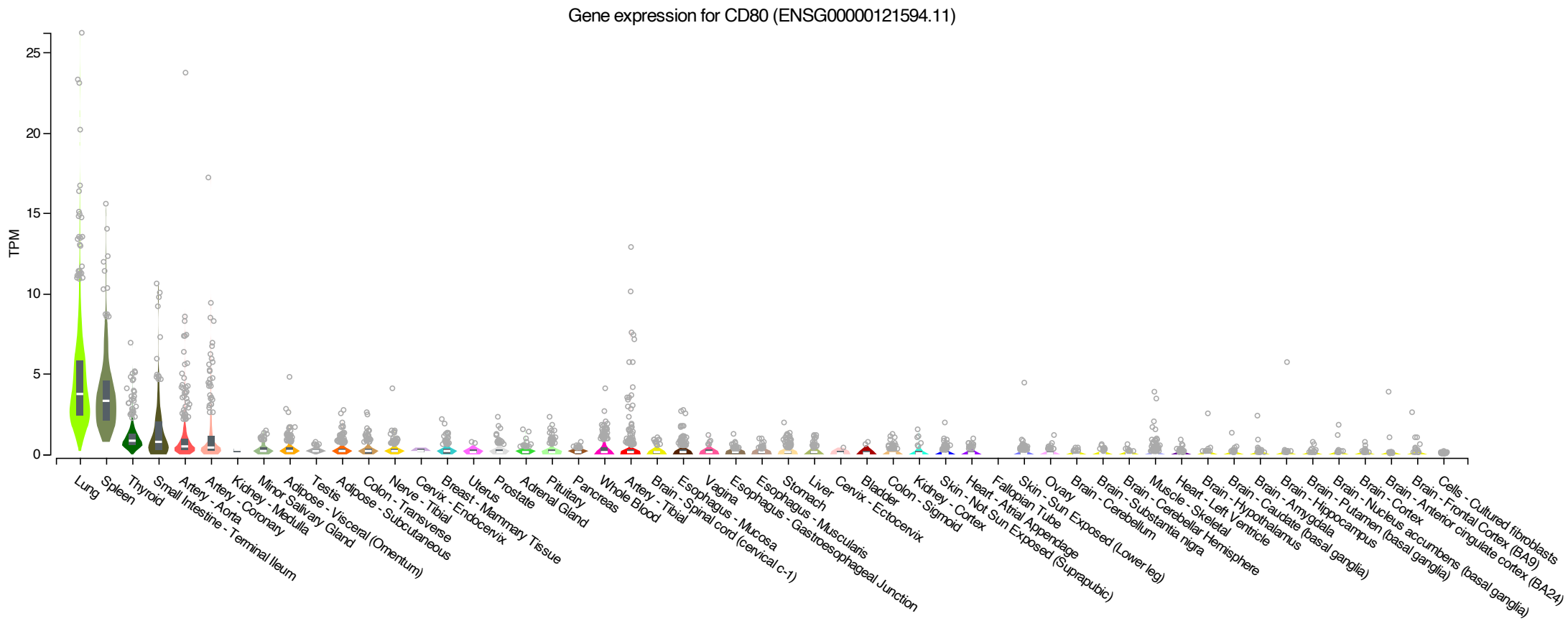
Gene expression for CD80 (ENSG00000121594.11)
TPM
0
5
10
15
20
25
Lung
Spleen
Thyroid
Small Intestine - Terminal Ileum
Artery - Aorta
Artery - Coronary
Kidney - Medulla
Minor Salivary Gland
Adipose - Visceral (Omentum)
Testis
Adipose - Subcutaneous
Colon - Transverse
Nerve - Tibial
Cervix - Endocervix
Breast - Mammary Tissue
Uterus
Prostate
Adrenal Gland
Pituitary
Pancreas
Whole Blood
Artery - Tibial
Brain - Spinal cord (cervical c-1)
Esophagus - Mucosa
Vagina
Esophagus - Muscularis
Esophagus - Gastroesophageal Junction
Stomach
Liver
Cervix - Ectocervix
Bladder
Colon - Sigmoid
Kidney - Cortex
Skin - Not Sun Exposed (Suprapubic)
Heart - Atrial Appendage
Fallopian Tube
Skin - Sun Exposed (Lower leg)
Ovary
Brain - Cerebellum
Brain - Substantia nigra
Brain - Cerebellar Hemisphere
Muscle - Skeletal
Heart - Left Ventricle
Brain - Hypothalamus
Brain - Amygdala
Brain - Caudate (basal ganglia)
Brain - Hippocampus
Brain - Putamen (basal ganglia)
Brain - Cortex
Brain - Nucleus accumbens (basal ganglia)
Brain - Anterior cingulate cortex (BA24)
Brain - Frontal Cortex (BA9)
Cells - Cultured fibroblasts

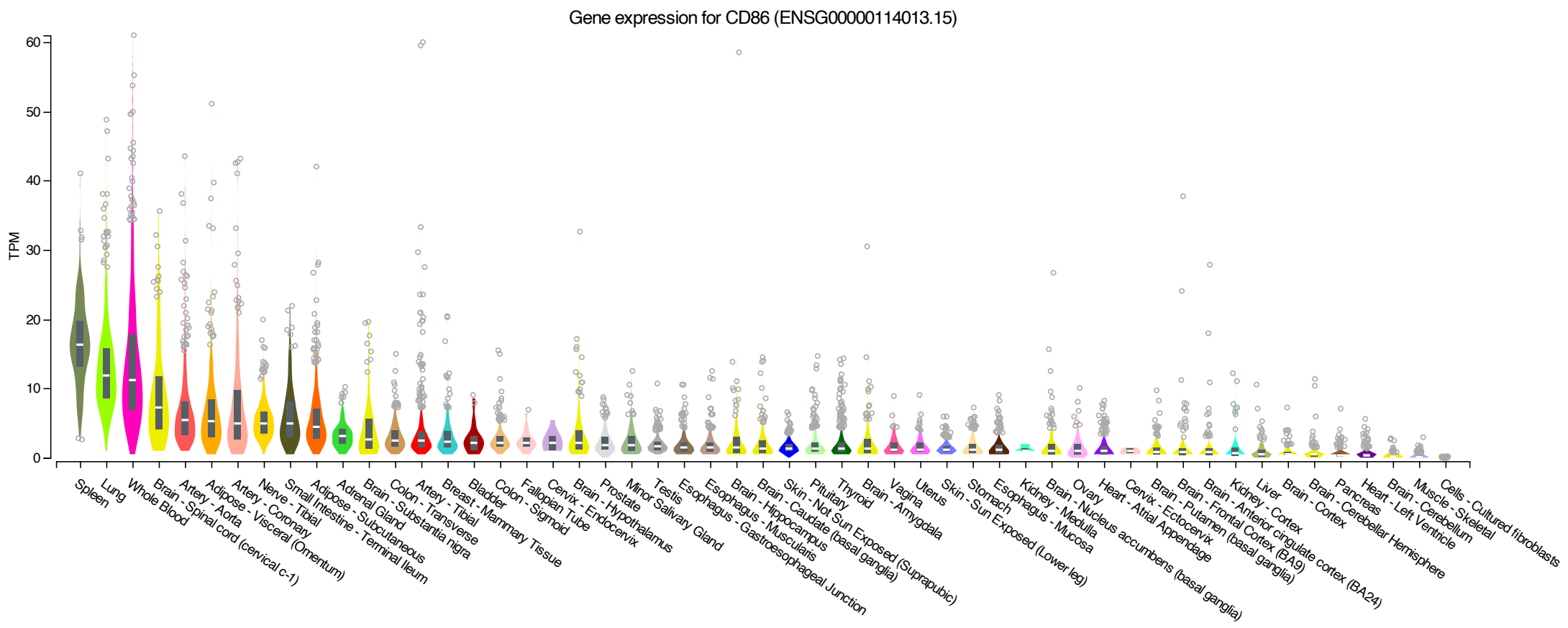
Gene expression for CD86 (ENSG00000114013.15)
TPM
0
10
20
30
40
50
60
Spleen
Lung
Whole Blood
Brain - Spinal cord (cervical c-1)
Artery - Aorta
Adipose - Visceral (Omentum)
Artery - Coronary
Nerve - Tibial
Small Intestine - Terminal Ileum
Adipose - Subcutaneous
Adrenal Gland
Brain - Substantia nigra
Colon - Transverse
Artery - Tibial
Breast - Mammary Tissue
Bladder
Colon - Sigmoid
Fallopian Tube
Cervix - Endocervix
Brain - Hypothalamus
Prostate
Minor Salivary Gland
Testis
Esophagus - Muscularis
Esophagus - Gastroesophageal Junction
Brain - Hippocampus
Brain - Caudate (basal ganglia)
Skin - Not Sun Exposed (Suprapubic)
Pituitary
Thyroid
Brain - Amygdala
Vagina
Uterus
Skin - Sun Exposed (Lower leg)
Stomach
Esophagus - Mucosa
Kidney - Medulla
Brain - Nucleus accumbens (basal ganglia)
Ovary
Heart - Atrial Appendage
Cervix - Ectocervix
Brain - Putamen (basal ganglia)
Brain - Frontal Cortex (BA9)
Brain - Anterior cingulate cortex (BA24)
Kidney - Cortex
Liver
Brain - Cortex
Brain - Cerebellar Hemisphere
Pancreas
Heart - Left Ventricle
Brain - Cerebellum
Muscle - Skeletal
Cells - Cultured fibroblasts

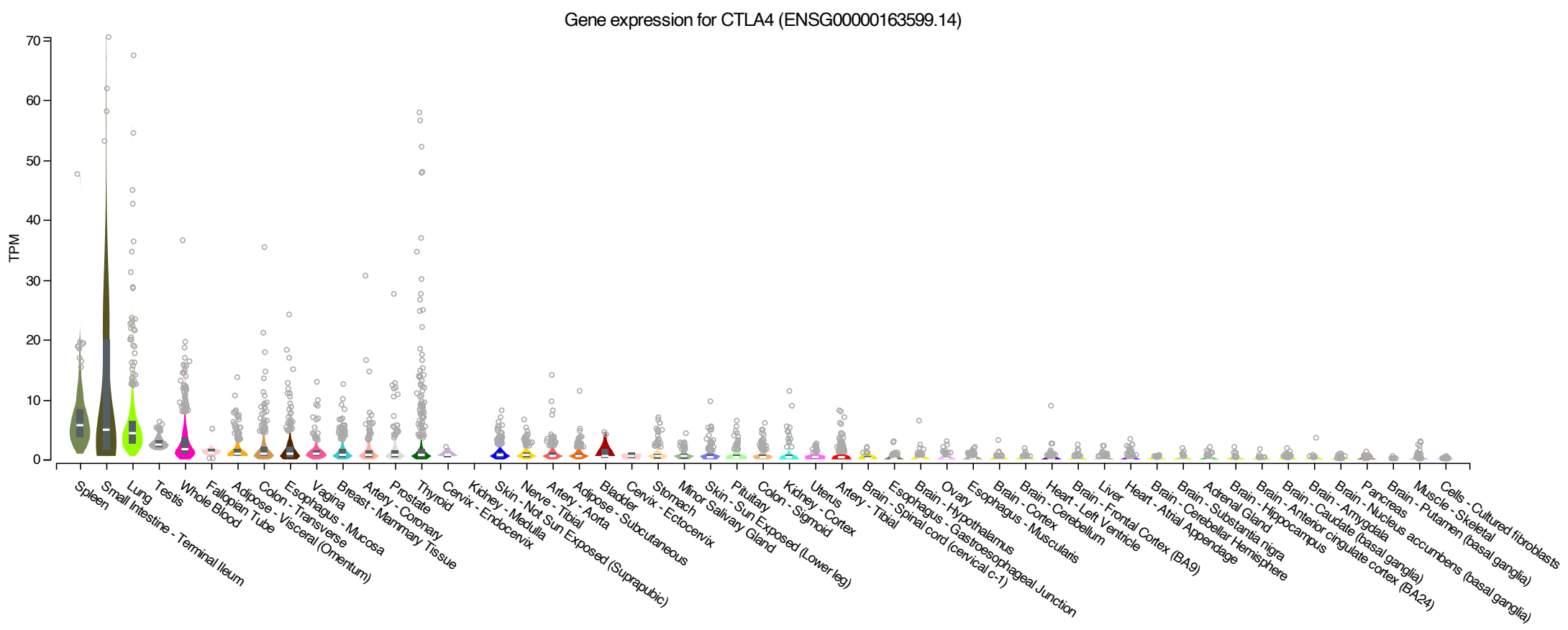
Gene expression for CTLA4 (ENSG00000163599.14)
TPM
0
10
20
30
40
50
60
70
Spleen
Small Intestine - Terminal Ileum
Lung
Testis
Whole Blood
Fallopian Tube
Adipose - Visceral (Omentum)
Colon - Transverse
Esophagus - Mucosa
Vagina
Breast - Mammary Tissue
Artery - Coronary
Prostate
Thyroid
Cervix - Endocervix
Kidney - Medulla
Skin - Not Sun Exposed (Suprapubic)
Nerve - Tibial
Artery - Aorta
Adipose - Subcutaneous
Bladder
Cervix - Ectocervix
Stomach
Minor Salivary Gland
Skin - Sun Exposed (Lower leg)
Pituitary
Colon - Sigmoid
Kidney - Cortex
Uterus
Artery - Tibial
Brain - Spinal cord (cervical c-1)
Esophagus - Gastroesophageal Junction
Brain - Hypothalamus
Ovary
Esophagus - Muscularis
Brain - Cortex
Brain - Cerebellum
Heart - Left Ventricle
Liver
Brain - Frontal Cortex (BA9)
Heart - Atrial Appendage
Brain - Cerebellar Hemisphere
Adrenal Gland
Brain - Substantia nigra
Brain - Hippocampus
Brain - Anterior cingulate cortex (BA24)
Brain - Caudate (basal ganglia)
Brain - Amygdala
Pancreas
Brain - Nucleus accumbens (basal ganglia)
Brain - Putamen (basal ganglia)
Muscle - Skeletal
Cells - Cultured fibroblasts

**Supplementary Table 1: List of biological processes associated to COVID-19 infection.**

| ID | Description | Size | NES | pvalue | p.adjust | ratio |
|---|---|---|---|---|---|---|
| GO:0016236 | macroautophagy | 295 | 2.152 | 1.96E-06 | 1.08E-04 | 0.432 |
| GO:0038093 | Fc receptor signaling pathway | 241 | 2.466 | 1.96E-06 | 1.08E-04 | 0.524 |
| GO:0002758 | innate immune response-activating signal transduction | 298 | 1.895 | 1.96E-06 | 1.08E-04 | 0.414 |
| GO:0015980 | energy derivation by oxidation of organic compounds | 285 | 2.125 | 1.96E-06 | 1.08E-04 | 0.330 |
| GO:0002440 | production of molecular mediator of immune response | 286 | 1.918 | 1.96E-06 | 1.08E-04 | 0.442 |
| GO:0072593 | reactive oxygen species metabolic process | 284 | 1.975 | 1.97E-06 | 1.08E-04 | 0.455 |
| GO:0019882 | antigen processing and presentation | 226 | 2.494 | 1.97E-06 | 1.08E-04 | 0.475 |
| GO:0006733 | oxidoreduction coenzyme metabolic process | 207 | 2.138 | 1.97E-06 | 1.08E-04 | 0.468 |
| GO:0002455 | humoral immune response mediated by circulating immunoglobulin | 150 | 3.043 | 1.97E-06 | 1.08E-04 | 0.739 |
| GO:0030100 | regulation of endocytosis | 281 | 1.998 | 1.97E-06 | 1.08E-04 | 0.477 |
| GO:0036294 | cellular response to decreased oxygen levels | 217 | 1.995 | 1.97E-06 | 1.08E-04 | 0.439 |
| GO:0046434 | organophosphate catabolic process | 248 | 2.072 | 1.97E-06 | 1.08E-04 | 0.351 |
| GO:0016052 | carbohydrate catabolic process | 199 | 2.129 | 1.97E-06 | 1.08E-04 | 0.393 |
| GO:0010324 | membrane invagination | 135 | 2.441 | 1.97E-06 | 1.08E-04 | 0.701 |
| GO:0022900 | electron transport chain | 186 | 2.139 | 1.97E-06 | 1.08E-04 | 0.577 |
| GO:0016054 | organic acid catabolic process | 275 | 1.950 | 1.97E-06 | 1.08E-04 | 0.465 |
| GO:0033209 | tumor necrosis factor-mediated signaling pathway | 167 | 1.998 | 1.97E-06 | 1.08E-04 | 0.463 |
| GO:0038094 | Fc-gamma receptor signaling pathway | 142 | 2.607 | 1.97E-06 | 1.08E-04 | 0.642 |
| GO:0002526 | acute inflammatory response | 220 | 2.600 | 1.97E-06 | 1.08E-04 | 0.558 |
| GO:1901136 | carbohydrate derivative catabolic process | 192 | 2.362 | 1.97E-06 | 1.08E-04 | 0.421 |
| GO:0061025 | membrane fusion | 154 | 2.299 | 1.97E-06 | 1.08E-04 | 0.505 |
| GO:0043281 | regulation of cysteine-type endopeptidase activity involved in apoptotic process | 215 | 1.926 | 1.97E-06 | 1.08E-04 | 0.400 |
| GO:0000070 | mitotic sister chromatid segregation | 151 | 2.028 | 1.97E-06 | 1.08E-04 | 0.273 |
| GO:0038061 | NIK/NF-kappaB signaling | 183 | 1.940 | 1.97E-06 | 1.08E-04 | 0.489 |
| GO:0050864 | regulation of B cell activation | 184 | 1.949 | 1.97E-06 | 1.08E-04 | 0.500 |
| GO:0007033 | vacuole organization | 163 | 2.177 | 1.97E-06 | 1.08E-04 | 0.459 |
| GO:0044106 | cellular amine metabolic process | 129 | 2.214 | 1.97E-06 | 1.08E-04 | 0.571 |
| GO:0055067 | monovalent inorganic cation homeostasis | 154 | 2.237 | 1.97E-06 | 1.08E-04 | 0.528 |
| GO:1902750 | negative regulation of cell cycle G2/M phase transition | 105 | 2.142 | 1.97E-06 | 1.08E-04 | 0.455 |
| GO:0002474 | antigen processing and presentation of peptide antigen via MHC class I | 96 | 2.509 | 1.97E-06 | 1.08E-04 | 0.505 |
| GO:0070498 | interleukin-1-mediated signaling pathway | 100 | 2.216 | 1.97E-06 | 1.08E-04 | 0.553 |
| GO:0031145 | anaphase-promoting complex-dependent catabolic process | 81 | 2.425 | 1.97E-06 | 1.08E-04 | 0.494 |
| GO:0009060 | aerobic respiration | 87 | 2.366 | 1.97E-06 | 1.08E-04 | 0.493 |
| GO:0032612 | interleukin-1 production | 115 | 2.113 | 1.97E-06 | 1.08E-04 | 0.557 |
| GO:0043648 | dicarboxylic acid metabolic process | 99 | 2.163 | 1.97E-06 | 1.08E-04 | 0.545 |
| GO:0061418 | response to hypoxia | 77 | 2.189 | 1.97E-06 | 1.08E-04 | 0.559 |
| GO:0045454 | cell redox homeostasis | 76 | 2.361 | 1.97E-06 | 1.08E-04 | 0.607 |
| GO:0045851 | pH reduction | 53 | 2.280 | 1.97E-06 | 1.08E-04 | 0.625 |
| GO:1901658 | glycosyl compound catabolic process | 43 | 2.264 | 1.98E-06 | 1.08E-04 | 0.457 |
| GO:0006614 | SRP-dependent cotranslational protein targeting to membrane | 105 | -2.113 | 2.03E-06 | 1.09E-04 | 0.611 |
| GO:0140014 | mitotic nuclear division | 264 | 1.766 | 3.93E-06 | 1.76E-04 | 0.227 |
| GO:0005996 | monosaccharide metabolic process | 292 | 1.833 | 3.93E-06 | 1.76E-04 | 0.384 |
| GO:0010466 | negative regulation of peptidase activity | 262 | 1.999 | 3.94E-06 | 1.76E-04 | 0.485 |
| GO:0008037 | cell recognition | 215 | 2.040 | 3.94E-06 | 1.76E-04 | 0.535 |
| GO:0016999 | antibiotic metabolic process | 151 | 2.091 | 3.94E-06 | 1.76E-04 | 0.480 |
| GO:0019682 | glyceraldehyde-3-phosphate metabolic process | 21 | 2.271 | 3.95E-06 | 1.76E-04 | 0.643 |
| GO:0006739 | NADP metabolic process | 32 | 2.271 | 3.95E-06 | 1.76E-04 | 0.542 |
| GO:0007040 | lysosome organization | 61 | 2.170 | 3.95E-06 | 1.76E-04 | 0.611 |
| GO:0031668 | cellular response to extracellular stimulus | 268 | 1.787 | 5.90E-06 | 2.50E-04 | 0.457 |
| GO:0009152 | purine ribonucleotide biosynthetic process | 280 | 1.768 | 5.90E-06 | 2.50E-04 | 0.303 |
| GO:0042180 | cellular ketone metabolic process | 248 | 1.838 | 5.90E-06 | 2.50E-04 | 0.481 |
| GO:0006376 | mRNA splice site selection | 53 | -2.321 | 6.08E-06 | 2.54E-04 | 0.323 |
| GO:0044839 | cell cycle G2/M phase transition | 266 | 1.740 | 7.85E-06 | 3.08E-04 | 0.379 |
| GO:0097164 | ammonium ion metabolic process | 205 | 1.884 | 7.87E-06 | 3.08E-04 | 0.472 |
| GO:0016051 | carbohydrate biosynthetic process | 214 | 1.864 | 7.87E-06 | 3.08E-04 | 0.444 |
| GO:0006900 | vesicle budding from membrane | 102 | 1.990 | 7.88E-06 | 3.08E-04 | 0.452 |
| GO:0000395 | mRNA 5'-splice site recognition | 29 | -2.217 | 8.11E-06 | 3.13E-04 | 0.583 |
| GO:1901988 | negative regulation of cell cycle phase transition | 267 | 1.757 | 9.82E-06 | 3.65E-04 | 0.432 |
| GO:0050701 | interleukin-1 secretion | 61 | 2.137 | 9.87E-06 | 3.65E-04 | 0.556 |
| GO:0034341 | response to interferon-gamma | 199 | 1.834 | 1.18E-05 | 4.15E-04 | 0.410 |
| GO:0007032 | endosome organization | 79 | 2.034 | 1.18E-05 | 4.15E-04 | 0.528 |
| GO:0006140 | regulation of nucleotide metabolic process | 146 | 1.924 | 1.38E-05 | 4.61E-04 | 0.368 |
| GO:0044242 | cellular lipid catabolic process | 217 | 1.835 | 1.38E-05 | 4.61E-04 | 0.456 |
| GO:0072350 | tricarboxylic acid metabolic process | 39 | 2.150 | 1.38E-05 | 4.61E-04 | 0.606 |
| GO:0032635 | interleukin-6 production | 161 | 1.859 | 1.57E-05 | 4.99E-04 | 0.505 |
| GO:1902850 | microtubule cytoskeleton organization involved in mitosis | 131 | 1.891 | 1.57E-05 | 4.99E-04 | 0.330 |
| GO:0033572 | transferrin transport | 36 | 2.163 | 1.58E-05 | 4.99E-04 | 0.643 |
| GO:0002532 | production of molecular mediator involved in inflammatory response | 72 | 2.091 | 1.58E-05 | 4.99E-04 | 0.608 |
| GO:0034381 | plasma lipoprotein particle clearance | 69 | 2.122 | 1.78E-05 | 5.44E-04 | 0.629 |
| GO:0048771 | tissue remodeling | 179 | 1.894 | 1.97E-05 | 5.85E-04 | 0.505 |
| GO:0051304 | chromosome separation | 90 | 1.965 | 1.97E-05 | 5.85E-04 | 0.267 |
| GO:0018196 | peptidyl-asparagine modification | 34 | 2.160 | 1.97E-05 | 5.85E-04 | 0.484 |
| GO:0009595 | detection of biotic stimulus | 23 | 2.171 | 1.98E-05 | 5.85E-04 | 0.800 |
| GO:0046466 | membrane lipid catabolic process | 35 | 2.157 | 2.76E-05 | 7.49E-04 | 0.692 |
| GO:0071216 | cellular response to biotic stimulus | 236 | 1.756 | 3.15E-05 | 8.42E-04 | 0.421 |
| GO:0050663 | cytokine secretion | 240 | 1.742 | 3.35E-05 | 8.87E-04 | 0.425 |
| GO:0042116 | macrophage activation | 95 | 1.970 | 3.95E-05 | 1.01E-03 | 0.529 |
| GO:0007584 | response to nutrient | 219 | 1.765 | 4.33E-05 | 1.09E-03 | 0.517 |
| GO:0000041 | transition metal ion transport | 112 | 1.924 | 4.34E-05 | 1.09E-03 | 0.467 |
| GO:0051262 | protein tetramerization | 172 | 1.812 | 4.53E-05 | 1.12E-03 | 0.311 |
| GO:0006364 | rRNA processing | 214 | -1.717 | 4.69E-05 | 1.13E-03 | 0.420 |
| GO:0060337 | type I interferon signaling pathway | 95 | 1.933 | 4.93E-05 | 1.16E-03 | 0.500 |
| GO:0006643 | membrane lipid metabolic process | 207 | 1.749 | 5.32E-05 | 1.24E-03 | 0.403 |
| GO:0019083 | viral transcription | 177 | -1.732 | 5.69E-05 | 1.32E-03 | 0.465 |
| GO:0042737 | drug catabolic process | 140 | 1.906 | 5.92E-05 | 1.37E-03 | 0.427 |
| GO:0000184 | decay | 120 | -1.808 | 6.51E-05 | 1.48E-03 | 0.576 |
| GO:0006081 | cellular aldehyde metabolic process | 74 | 2.030 | 6.52E-05 | 1.48E-03 | 0.568 |
| GO:0010499 | proteasomal ubiquitin-independent protein catabolic process | 23 | 2.118 | 6.70E-05 | 1.52E-03 | 0.524 |
| GO:0051701 | interaction with host | 209 | 1.707 | 7.47E-05 | 1.64E-03 | 0.416 |
| GO:0002220 | pathway | 116 | 1.835 | 8.07E-05 | 1.74E-03 | 0.580 |
| GO:0051225 | spindle assembly | 108 | 1.833 | 8.87E-05 | 1.87E-03 | 0.284 |
| GO:0050764 | regulation of phagocytosis | 98 | 1.880 | 8.88E-05 | 1.87E-03 | 0.547 |
| GO:0032418 | lysosome localization | 74 | 1.947 | 9.28E-05 | 1.93E-03 | 0.632 |
| GO:0070646 | protein modification by small protein removal | 299 | 1.607 | 1.04E-04 | 2.16E-03 | 0.440 |
| GO:0006479 | protein methylation | 179 | -1.744 | 1.12E-04 | 2.28E-03 | 0.421 |
| GO:0016197 | endosomal transport | 224 | 1.670 | 1.12E-04 | 2.28E-03 | 0.415 |
| GO:0090288 | negative regulation of cellular response to growth factor stimulus | 166 | -1.858 | 1.18E-04 | 2.38E-03 | 0.415 |
| GO:0043112 | receptor metabolic process | 192 | 1.733 | 1.24E-04 | 2.49E-03 | 0.455 |
| GO:0032868 | response to insulin | 272 | 1.641 | 1.51E-04 | 2.93E-03 | 0.402 |
| GO:1905330 | regulation of morphogenesis of an epithelium | 180 | 1.752 | 1.77E-04 | 3.35E-03 | 0.496 |
| GO:0046365 | monosaccharide catabolic process | 67 | 1.941 | 1.80E-04 | 3.39E-03 | 0.408 |
| GO:0097205 | renal filtration | 22 | 2.059 | 1.83E-04 | 3.45E-03 | 0.615 |
| GO:0090175 | regulation of establishment of planar polarity | 110 | 1.799 | 2.21E-04 | 4.05E-03 | 0.512 |
| GO:0032722 | positive regulation of chemokine production | 58 | 1.937 | 2.45E-04 | 4.42E-03 | 0.561 |
| GO:1902600 | proton transmembrane transport | 163 | 1.771 | 2.52E-04 | 4.52E-03 | 0.547 |
| GO:0030193 | regulation of blood coagulation | 79 | 1.915 | 2.62E-04 | 4.67E-03 | 0.511 |
| GO:0006890 | retrograde vesicle-mediated transport, Golgi to ER | 86 | 1.818 | 2.66E-04 | 4.72E-03 | 0.455 |
| GO:0002573 | myeloid leukocyte differentiation | 204 | 1.656 | 2.85E-04 | 4.97E-03 | 0.497 |
| GO:0002793 | positive regulation of peptide secretion | 288 | 1.608 | 2.97E-04 | 5.13E-03 | 0.361 |
| GO:0060759 | regulation of response to cytokine stimulus | 190 | 1.678 | 3.05E-04 | 5.25E-03 | 0.430 |
| GO:1901568 | fatty acid derivative metabolic process | 167 | 1.764 | 3.11E-04 | 5.32E-03 | 0.355 |
| GO:0042133 | neurotransmitter metabolic process | 153 | 1.777 | 3.11E-04 | 5.32E-03 | 0.494 |
| GO:0031638 | zymogen activation | 53 | 1.988 | 3.12E-04 | 5.32E-03 | 0.567 |
| GO:0036109 | alpha-linolenic acid metabolic process | 13 | 2.011 | 3.40E-04 | 5.74E-03 | 0.727 |
| GO:0034976 | response to endoplasmic reticulum stress | 285 | 1.564 | 3.59E-04 | 5.95E-03 | 0.416 |
| GO:0031639 | plasminogen activation | 25 | 2.011 | 3.73E-04 | 6.09E-03 | 0.538 |
| GO:0046364 | monosaccharide biosynthetic process | 98 | 1.813 | 3.77E-04 | 6.14E-03 | 0.443 |
| GO:1902476 | chloride transmembrane transport | 88 | 1.956 | 3.93E-04 | 6.31E-03 | 0.424 |
| GO:0032637 | interleukin-8 production | 82 | 1.841 | 3.97E-04 | 6.35E-03 | 0.533 |
| GO:0009566 | fertilization | 182 | 1.803 | 4.20E-04 | 6.69E-03 | 0.451 |
| GO:1902036 | regulation of hematopoietic stem cell differentiation | 72 | 1.812 | 4.73E-04 | 7.48E-03 | 0.379 |
| GO:0006801 | superoxide metabolic process | 73 | 1.860 | 4.92E-04 | 7.66E-03 | 0.588 |
| GO:0006575 | cellular modified amino acid metabolic process | 202 | 1.666 | 4.92E-04 | 7.66E-03 | 0.484 |
| GO:0043174 | nucleoside salvage | 16 | 1.991 | 4.93E-04 | 7.66E-03 | 0.462 |
| GO:0006027 | glycosaminoglycan catabolic process | 61 | 1.949 | 5.19E-04 | 7.99E-03 | 0.586 |
| GO:1903008 | organelle disassembly | 101 | 1.730 | 5.54E-04 | 8.49E-03 | 0.467 |
| GO:0090092 | signaling pathway | 241 | -1.668 | 5.73E-04 | 8.77E-03 | 0.366 |
| GO:0008217 | regulation of blood pressure | 182 | 1.756 | 5.80E-04 | 8.84E-03 | 0.325 |
| GO:0006888 | ER to Golgi vesicle-mediated transport | 212 | 1.585 | 5.96E-04 | 9.05E-03 | 0.420 |
| GO:0043154 | in apoptotic process | 84 | 1.802 | 6.10E-04 | 9.20E-03 | 0.500 |
| GO:2000379 | positive regulation of reactive oxygen species metabolic process | 102 | 1.768 | 6.32E-04 | 9.50E-03 | 0.542 |
| GO:0045730 | respiratory burst | 37 | 1.930 | 6.46E-04 | 9.65E-03 | 0.552 |
| GO:0007029 | endoplasmic reticulum organization | 57 | 1.873 | 6.47E-04 | 9.65E-03 | 0.575 |
| GO:0071772 | response to BMP | 170 | -1.754 | 6.67E-04 | 9.80E-03 | 0.403 |
| GO:0002755 | MyD88-dependent toll-like receptor signaling pathway | 36 | 1.901 | 6.84E-04 | 9.94E-03 | 0.471 |
| GO:1903557 | production | 88 | 1.760 | 7.18E-04 | 1.03E-02 | 0.486 |
| GO:0006672 | ceramide metabolic process | 92 | 1.780 | 7.34E-04 | 1.04E-02 | 0.530 |
| GO:1903034 | regulation of response to wounding | 179 | 1.671 | 7.51E-04 | 1.06E-02 | 0.448 |
| GO:0007034 | vacuolar transport | 142 | 1.629 | 7.74E-04 | 1.08E-02 | 0.414 |
| GO:0090263 | positive regulation of canonical Wnt signaling pathway | 147 | 1.666 | 7.79E-04 | 1.08E-02 | 0.327 |
| GO:1901216 | positive regulation of neuron death | 94 | 1.742 | 7.81E-04 | 1.08E-02 | 0.533 |
| GO:0048260 | positive regulation of receptor-mediated endocytosis | 51 | 1.895 | 7.82E-04 | 1.08E-02 | 0.576 |
| GO:0033865 | nucleoside bisphosphate metabolic process | 140 | 1.670 | 8.09E-04 | 1.11E-02 | 0.346 |
| GO:1901605 | alpha-amino acid metabolic process | 222 | 1.612 | 8.46E-04 | 1.15E-02 | 0.449 |
| GO:0010821 | regulation of mitochondrion organization | 182 | 1.587 | 8.53E-04 | 1.16E-02 | 0.290 |
| GO:0016125 | sterol metabolic process | 166 | 1.657 | 8.81E-04 | 1.19E-02 | 0.352 |
| GO:2001234 | negative regulation of apoptotic signaling pathway | 230 | 1.568 | 8.97E-04 | 1.21E-02 | 0.395 |
| GO:0098754 | detoxification | 131 | 1.717 | 9.41E-04 | 1.25E-02 | 0.506 |
| GO:0006919 | apoptotic process | 86 | 1.753 | 9.63E-04 | 1.27E-02 | 0.426 |
| GO:0046470 | phosphatidylcholine metabolic process | 83 | 1.804 | 9.73E-04 | 1.28E-02 | 0.481 |
| GO:0051321 | meiotic cell cycle | 249 | 1.596 | 9.94E-04 | 1.30E-02 | 0.275 |
| GO:0097352 | autophagosome maturation | 38 | 1.873 | 1.02E-03 | 1.33E-02 | 0.424 |
| GO:0001510 | RNA methylation | 81 | -1.737 | 1.17E-03 | 1.48E-02 | 0.279 |
| GO:0032799 | low-density lipoprotein receptor particle metabolic process | 23 | 1.916 | 1.21E-03 | 1.52E-02 | 0.684 |
| GO:0045471 | response to ethanol | 125 | 1.733 | 1.22E-03 | 1.53E-02 | 0.471 |
| GO:0031060 | regulation of histone methylation | 65 | -1.774 | 1.23E-03 | 1.53E-02 | 0.509 |
| GO:1903531 | negative regulation of secretion by cell | 211 | 1.615 | 1.28E-03 | 1.57E-02 | 0.410 |
| GO:0019321 | pentose metabolic process | 15 | 1.912 | 1.30E-03 | 1.59E-02 | 0.500 |
| GO:0048259 | regulation of receptor-mediated endocytosis | 103 | 1.743 | 1.30E-03 | 1.60E-02 | 0.452 |
| GO:0061647 | histone H3-K9 modification | 48 | -1.819 | 1.38E-03 | 1.67E-02 | 0.564 |
| GO:0072329 | monocarboxylic acid catabolic process | 132 | 1.650 | 1.47E-03 | 1.77E-02 | 0.411 |
| GO:0017001 | antibiotic catabolic process | 58 | 1.825 | 1.48E-03 | 1.78E-02 | 0.500 |
| GO:0030041 | actin filament polymerization | 182 | 1.576 | 1.56E-03 | 1.86E-02 | 0.493 |
| GO:0042053 | regulation of dopamine metabolic process | 20 | 1.901 | 1.56E-03 | 1.86E-02 | 0.818 |
| GO:0042058 | regulation of epidermal growth factor receptor signaling pathway | 86 | 1.714 | 1.59E-03 | 1.88E-02 | 0.576 |
| GO:0140029 | exocytic process | 80 | 1.732 | 1.70E-03 | 1.98E-02 | 0.526 |
| GO:0051984 | positive regulation of chromosome segregation | 28 | 1.864 | 1.75E-03 | 2.02E-02 | 0.240 |
| GO:0061640 | cytoskeleton-dependent cytokinesis | 100 | 1.688 | 1.76E-03 | 2.03E-02 | 0.417 |
| GO:0006123 | mitochondrial electron transport, cytochrome c to oxygen | 21 | 1.890 | 1.77E-03 | 2.03E-02 | 0.706 |
| GO:1904668 | positive regulation of ubiquitin protein ligase activity | 12 | 1.889 | 1.79E-03 | 2.05E-02 | 0.545 |
| GO:0042471 | ear morphogenesis | 118 | -1.775 | 1.84E-03 | 2.10E-02 | 0.409 |
| GO:0031146 | process | 95 | 1.658 | 1.86E-03 | 2.12E-02 | 0.373 |
| GO:0033077 | T cell differentiation in thymus | 70 | -1.737 | 1.87E-03 | 2.12E-02 | 0.455 |
| GO:1903509 | liposaccharide metabolic process | 111 | 1.654 | 1.89E-03 | 2.14E-02 | 0.405 |
| GO:0002576 | platelet degranulation | 128 | 1.636 | 1.90E-03 | 2.15E-02 | 0.516 |
| GO:0050862 | positive regulation of T cell receptor signaling pathway | 14 | -1.897 | 2.02E-03 | 2.25E-02 | 0.750 |
| GO:0050820 | positive regulation of coagulation | 27 | 1.872 | 2.07E-03 | 2.31E-02 | 0.579 |
| GO:0043277 | apoptotic cell clearance | 46 | 1.813 | 2.12E-03 | 2.34E-02 | 0.750 |
| GO:0022618 | ribonucleoprotein complex assembly | 277 | -1.471 | 2.14E-03 | 2.36E-02 | 0.294 |
| GO:0045410 | positive regulation of interleukin-6 biosynthetic process | 15 | 1.878 | 2.17E-03 | 2.39E-02 | 0.733 |
| GO:0036258 | multivesicular body assembly | 30 | 1.828 | 2.19E-03 | 2.40E-02 | 0.607 |
| GO:0010823 | negative regulation of mitochondrion organization | 55 | 1.758 | 2.20E-03 | 2.41E-02 | 0.409 |
| GO:0036124 | histone H3-K9 trimethylation | 16 | -1.887 | 2.23E-03 | 2.43E-02 | 0.615 |
| GO:0006907 | pinocytosis | 21 | 1.874 | 2.25E-03 | 2.44E-02 | 0.563 |
| GO:0090148 | membrane fission | 12 | 1.869 | 2.26E-03 | 2.45E-02 | 0.700 |
| GO:0007229 | integrin-mediated signaling pathway | 103 | 1.659 | 2.34E-03 | 2.52E-02 | 0.459 |
| GO:0010257 | NADH dehydrogenase complex assembly | 64 | 1.698 | 2.35E-03 | 2.52E-02 | 0.383 |
| GO:1900407 | regulation of cellular response to oxidative stress | 88 | 1.700 | 2.38E-03 | 2.54E-02 | 0.441 |
| GO:0032506 | cytokinetic process | 39 | 1.779 | 2.42E-03 | 2.58E-02 | 0.459 |
| GO:0001845 | phagolysosome assembly | 13 | 1.863 | 2.43E-03 | 2.58E-02 | 0.600 |
| GO:1904385 | cellular response to angiotensin | 25 | 1.867 | 2.45E-03 | 2.59E-02 | 0.400 |
| GO:0045132 | meiotic chromosome segregation | 90 | 1.738 | 2.48E-03 | 2.60E-02 | 0.234 |
| GO:0060395 | SMAD protein signal transduction | 70 | -1.787 | 2.51E-03 | 2.63E-02 | 0.412 |
| GO:0021700 | developmental maturation | 284 | 1.515 | 2.56E-03 | 2.66E-02 | 0.442 |
| GO:0015718 | monocarboxylic acid transport | 162 | 1.632 | 2.64E-03 | 2.72E-02 | 0.537 |
| GO:0090596 | sensory organ morphogenesis | 256 | -1.590 | 2.65E-03 | 2.72E-02 | 0.419 |
| GO:0043094 | cellular metabolic compound salvage | 34 | 1.821 | 2.65E-03 | 2.72E-02 | 0.385 |
| GO:0072523 | purine-containing compound catabolic process | 51 | 1.772 | 2.66E-03 | 2.73E-02 | 0.278 |
| GO:0089718 | amino acid import across plasma membrane | 23 | 1.860 | 2.68E-03 | 2.74E-02 | 0.750 |
| GO:0033194 | response to hydroperoxide | 20 | 1.854 | 2.68E-03 | 2.74E-02 | 0.647 |
| GO:0017157 | regulation of exocytosis | 217 | 1.548 | 2.73E-03 | 2.77E-02 | 0.394 |
| GO:0030510 | regulation of BMP signaling pathway | 91 | -1.744 | 2.76E-03 | 2.79E-02 | 0.381 |
| GO:0043627 | response to estrogen | 73 | 1.734 | 2.87E-03 | 2.89E-02 | 0.477 |
| GO:0003071 | pressure | 25 | 1.849 | 2.87E-03 | 2.89E-02 | 0.600 |
| GO:1903076 | regulation of protein localization to plasma membrane | 95 | 1.664 | 2.96E-03 | 2.96E-02 | 0.477 |
| GO:0097066 | response to thyroid hormone | 26 | 1.841 | 2.99E-03 | 2.98E-02 | 0.556 |
| GO:0043300 | regulation of leukocyte degranulation | 46 | 1.742 | 3.06E-03 | 3.04E-02 | 0.550 |
| GO:2000482 | regulation of interleukin-8 secretion | 25 | 1.828 | 3.06E-03 | 3.04E-02 | 0.700 |
| GO:0002082 | regulation of oxidative phosphorylation | 31 | 1.824 | 3.08E-03 | 3.05E-02 | 0.455 |
| GO:0043624 | cellular protein complex disassembly | 217 | 1.483 | 3.10E-03 | 3.07E-02 | 0.447 |
| GO:0098656 | anion transmembrane transport | 288 | 1.497 | 3.20E-03 | 3.16E-02 | 0.469 |
| GO:0019674 | NAD metabolic process | 76 | 1.687 | 3.25E-03 | 3.20E-02 | 0.519 |
| GO:0046835 | carbohydrate phosphorylation | 24 | 1.832 | 3.26E-03 | 3.20E-02 | 0.556 |
| GO:2001057 | reactive nitrogen species metabolic process | 85 | 1.681 | 3.30E-03 | 3.23E-02 | 0.509 |
| GO:0046683 | response to organophosphorus | 134 | 1.611 | 3.36E-03 | 3.28E-02 | 0.354 |
| GO:0009451 | RNA modification | 163 | -1.523 | 3.39E-03 | 3.29E-02 | 0.264 |
| GO:0070125 | mitochondrial translational elongation | 88 | 1.613 | 3.43E-03 | 3.33E-02 | 0.481 |
| GO:0051650 | establishment of vesicle localization | 290 | 1.454 | 3.54E-03 | 3.42E-02 | 0.429 |
| GO:0031348 | negative regulation of defense response | 239 | 1.504 | 3.61E-03 | 3.47E-02 | 0.388 |
| GO:0018023 | peptidyl-lysine trimethylation | 44 | -1.751 | 3.61E-03 | 3.47E-02 | 0.500 |
| GO:0006767 | water-soluble vitamin metabolic process | 88 | 1.633 | 3.68E-03 | 3.53E-02 | 0.449 |
| GO:0006836 | neurotransmitter transport | 269 | 1.524 | 3.69E-03 | 3.53E-02 | 0.426 |
| GO:0010507 | negative regulation of autophagy | 84 | 1.640 | 3.87E-03 | 3.67E-02 | 0.438 |
| GO:0051307 | meiotic chromosome separation | 25 | 1.825 | 3.92E-03 | 3.71E-02 | 0.357 |
| GO:0060986 | endocrine hormone secretion | 47 | -1.818 | 3.96E-03 | 3.73E-02 | 0.333 |
| GO:0032928 | regulation of superoxide anion generation | 22 | 1.814 | 4.04E-03 | 3.79E-02 | 0.824 |
| GO:0034446 | substrate adhesion-dependent cell spreading | 100 | 1.639 | 4.05E-03 | 3.79E-02 | 0.508 |
| GO:0060042 | retina morphogenesis in camera-type eye | 49 | -1.819 | 4.13E-03 | 3.84E-02 | 0.412 |
| GO:0072665 | protein localization to vacuole | 65 | 1.657 | 4.21E-03 | 3.91E-02 | 0.536 |
| GO:0008347 | glial cell migration | 49 | 1.738 | 4.39E-03 | 4.04E-02 | 0.455 |
| GO:0051931 | regulation of sensory perception | 40 | 1.815 | 4.39E-03 | 4.04E-02 | 0.583 |
| GO:0097028 | dendritic cell differentiation | 42 | 1.741 | 4.47E-03 | 4.08E-02 | 0.531 |
| GO:0006607 | NLS-bearing protein import into nucleus | 18 | -1.804 | 4.54E-03 | 4.14E-02 | 0.389 |
| GO:0016573 | histone acetylation | 156 | -1.507 | 4.58E-03 | 4.17E-02 | 0.386 |
| GO:0060021 | roof of mouth development | 89 | -1.674 | 4.59E-03 | 4.17E-02 | 0.313 |
| GO:0006730 | one-carbon metabolic process | 27 | 1.778 | 4.63E-03 | 4.20E-02 | 0.609 |
| GO:0051187 | cofactor catabolic process | 65 | 1.726 | 4.66E-03 | 4.21E-02 | 0.324 |
| GO:0051785 | positive regulation of nuclear division | 66 | 1.696 | 4.72E-03 | 4.27E-02 | 0.317 |
| GO:0043032 | positive regulation of macrophage activation | 27 | 1.803 | 4.75E-03 | 4.28E-02 | 0.533 |
| GO:0034142 | toll-like receptor 4 signaling pathway | 35 | 1.765 | 4.77E-03 | 4.28E-02 | 0.720 |
| GO:0006691 | leukotriene metabolic process | 31 | 1.805 | 4.79E-03 | 4.30E-02 | 0.429 |
| GO:0044275 | cellular carbohydrate catabolic process | 45 | 1.722 | 4.84E-03 | 4.32E-02 | 0.441 |
| GO:0030316 | osteoclast differentiation | 97 | 1.599 | 4.86E-03 | 4.33E-02 | 0.542 |
| GO:0008637 | apoptotic mitochondrial changes | 124 | 1.548 | 4.90E-03 | 4.35E-02 | 0.367 |
| GO:0051881 | regulation of mitochondrial membrane potential | 72 | 1.651 | 4.91E-03 | 4.35E-02 | 0.434 |
| GO:0051965 | positive regulation of synapse assembly | 68 | -1.788 | 4.93E-03 | 4.36E-02 | 0.650 |
| GO:0051988 | regulation of attachment of spindle microtubules to kinetochore | 12 | 1.803 | 5.05E-03 | 4.44E-02 | 0.750 |
| GO:0031056 | regulation of histone modification | 143 | -1.518 | 5.06E-03 | 4.44E-02 | 0.387 |
| GO:0043902 | positive regulation of multi-organism process | 188 | 1.475 | 5.16E-03 | 4.53E-02 | 0.376 |
| GO:0050729 | positive regulation of inflammatory response | 153 | 1.549 | 5.19E-03 | 4.53E-02 | 0.389 |
| GO:0097191 | extrinsic apoptotic signaling pathway | 224 | 1.456 | 5.25E-03 | 4.58E-02 | 0.405 |
| GO:0001774 | microglial cell activation | 48 | 1.705 | 5.30E-03 | 4.60E-02 | 0.417 |
| GO:0034058 | endosomal vesicle fusion | 10 | 1.791 | 5.36E-03 | 4.63E-02 | 0.700 |
| GO:0035635 | entry of bacterium into host cell | 15 | 1.791 | 5.36E-03 | 4.63E-02 | 0.600 |
| GO:0002385 | mucosal immune response | 36 | -1.784 | 5.43E-03 | 4.68E-02 | 0.444 |
| GO:0006022 | aminoglycan metabolic process | 170 | 1.539 | 5.51E-03 | 4.71E-02 | 0.402 |
| GO:2000756 | regulation of peptidyl-lysine acetylation | 59 | -1.657 | 5.67E-03 | 4.84E-02 | 0.383 |
| GO:1905819 | negative regulation of chromosome separation | 40 | 1.704 | 5.76E-03 | 4.89E-02 | 0.324 |
| GO:0019432 | triglyceride biosynthetic process | 41 | 1.747 | 5.84E-03 | 4.91E-02 | 0.500 |
| GO:0001659 | temperature homeostasis | 173 | 1.508 | 5.93E-03 | 4.96E-02 | 0.316 |

**Legend**

| | |
|---|---|
| **Size** | Number of genes defining the biological process in the GO database |
| **NES** | its directionality |
| **pvalue** | permutations) |
| **p.adjust** | FDR-corrected p-value |
| **ratio*** | biological process |

msigdb.org/gsea/doc/GSEAUserGuideFrame.html

**Supplementary Table 2: List of biological processes associated to treatment with abatacept**

| ID | Description | Size | NES | pvalue | p.adjust | ratio |
|---|---|---|---|---|---|---|
| GO:0140014 | mitotic nuclear division | 264 | -2.222 | 1.75E-06 | 1.28E-04 | 0.394 |
| GO:0038093 | Fc receptor signaling pathway | 241 | -3.021 | 1.76E-06 | 1.28E-04 | 0.371 |
| GO:1901988 | negative regulation of cell cycle phase transition | 267 | -1.811 | 1.76E-06 | 1.28E-04 | 0.295 |
| GO:0002440 | production of molecular mediator of immune response | 286 | -2.758 | 1.76E-06 | 1.28E-04 | 0.216 |
| GO:0045637 | regulation of myeloid cell differentiation | 251 | -1.959 | 1.76E-06 | 1.28E-04 | 0.388 |
| GO:0019882 | antigen processing and presentation | 226 | -1.972 | 1.77E-06 | 1.28E-04 | 0.369 |
| GO:0050864 | regulation of B cell activation | 184 | -2.765 | 1.81E-06 | 1.28E-04 | 0.326 |
| GO:0002526 | acute inflammatory response | 220 | -3.092 | 1.81E-06 | 1.28E-04 | 0.379 |
| GO:0031497 | chromatin assembly | 165 | -2.643 | 1.81E-06 | 1.28E-04 | 0.401 |
| GO:0000070 | mitotic sister chromatid segregation | 151 | -2.243 | 1.81E-06 | 1.28E-04 | 0.382 |
| GO:2001251 | negative regulation of chromosome organization | 146 | -2.297 | 1.82E-06 | 1.28E-04 | 0.380 |
| GO:0038094 | Fc-gamma receptor signaling pathway | 142 | -3.182 | 1.82E-06 | 1.28E-04 | 0.531 |
| GO:0008037 | cell recognition | 215 | -2.611 | 1.82E-06 | 1.28E-04 | 0.328 |
| GO:0045814 | negative regulation of gene expression, epigenetic | 136 | -2.335 | 1.83E-06 | 1.28E-04 | 0.441 |
| GO:1902850 | microtubule cytoskeleton organization involved in mitosis | 131 | -2.213 | 1.83E-06 | 1.28E-04 | 0.461 |
| GO:0002455 | humoral immune response mediated by circulating immunoglobulin | 150 | -3.253 | 1.85E-06 | 1.28E-04 | 0.612 |
| GO:0010324 | membrane invagination | 135 | -2.837 | 1.85E-06 | 1.28E-04 | 0.418 |
| GO:0051225 | spindle assembly | 108 | -2.041 | 1.85E-06 | 1.28E-04 | 0.443 |
| GO:0045652 | regulation of megakaryocyte differentiation | 79 | -2.168 | 1.87E-06 | 1.28E-04 | 0.432 |
| GO:0051304 | chromosome separation | 90 | -2.215 | 1.87E-06 | 1.28E-04 | 0.297 |
| GO:0006335 | DNA replication-dependent nucleosome assembly | 32 | -2.217 | 1.94E-06 | 1.28E-04 | 0.531 |
| GO:0019731 | antibacterial humoral response | 46 | -2.234 | 1.95E-06 | 1.28E-04 | 0.655 |
| GO:0042274 | ribosomal small subunit biogenesis | 67 | 2.163 | 2.13E-06 | 1.28E-04 | 0.569 |
| GO:0042273 | ribosomal large subunit biogenesis | 71 | 2.214 | 2.14E-06 | 1.28E-04 | 0.500 |
| GO:0002181 | cytoplasmic translation | 100 | 2.017 | 2.18E-06 | 1.28E-04 | 0.404 |
| GO:0006614 | SRP-dependent cotranslational protein targeting to membrane | 105 | 2.746 | 2.18E-06 | 1.28E-04 | 0.704 |
| GO:0000184 | nuclear-transcribed mRNA catabolic process, nonsense-mediated decay | 120 | 2.663 | 2.21E-06 | 1.28E-04 | 0.563 |
| GO:0019083 | viral transcription | 177 | 2.439 | 2.27E-06 | 1.28E-04 | 0.494 |
| GO:0006413 | translational initiation | 193 | 2.229 | 2.28E-06 | 1.28E-04 | 0.421 |
| GO:0006364 | rRNA processing | 214 | 2.183 | 2.31E-06 | 1.28E-04 | 0.419 |
| GO:0051290 | protein heterotetramerization | 54 | -2.125 | 3.82E-06 | 1.92E-04 | 0.396 |
| GO:1905819 | negative regulation of chromosome separation | 40 | -2.197 | 3.87E-06 | 1.92E-04 | 0.382 |
| GO:0007093 | mitotic cell cycle checkpoint | 165 | -1.849 | 5.41E-06 | 2.58E-04 | 0.300 |
| GO:0060147 | regulation of posttranscriptional gene silencing | 117 | -2.000 | 5.51E-06 | 2.58E-04 | 0.306 |
| GO:0009451 | RNA modification | 163 | 1.829 | 6.75E-06 | 2.97E-04 | 0.294 |
| GO:0038111 | interleukin-7-mediated signaling pathway | 30 | -2.135 | 7.81E-06 | 3.37E-04 | 0.571 |
| GO:0000028 | ribosomal small subunit assembly | 19 | 2.131 | 8.03E-06 | 3.44E-04 | 0.706 |
| GO:0019730 | antimicrobial humoral response | 122 | -2.052 | 1.13E-05 | 4.74E-04 | 0.450 |
| GO:0090224 | regulation of spindle organization | 41 | -2.110 | 1.35E-05 | 5.51E-04 | 0.568 |
| GO:0030261 | chromosome condensation | 47 | -2.099 | 1.35E-05 | 5.51E-04 | 0.667 |
| GO:0000027 | ribosomal large subunit assembly | 30 | 2.124 | 1.64E-05 | 6.33E-04 | 0.679 |
| GO:0001510 | RNA methylation | 81 | 1.953 | 2.15E-05 | 8.25E-04 | 0.359 |
| GO:0006890 | retrograde vesicle-mediated transport, Golgi to ER | 86 | -1.947 | 2.99E-05 | 1.13E-03 | 0.418 |
| GO:0008033 | tRNA processing | 130 | 1.803 | 3.11E-05 | 1.16E-03 | 0.250 |
| GO:0006338 | chromatin remodeling | 182 | -1.771 | 3.23E-05 | 1.19E-03 | 0.369 |
| GO:0090068 | positive regulation of cell cycle process | 298 | -1.670 | 3.33E-05 | 1.22E-03 | 0.295 |
| GO:0051310 | metaphase plate congression | 57 | -2.011 | 3.43E-05 | 1.25E-03 | 0.431 |
| GO:1901216 | positive regulation of neuron death | 94 | -1.930 | 4.30E-05 | 1.54E-03 | 0.382 |
| GO:0070125 | mitochondrial translational elongation | 88 | 1.892 | 4.55E-05 | 1.60E-03 | 0.322 |
| GO:0051321 | meiotic cell cycle | 249 | -1.721 | 1.06E-04 | 3.60E-03 | 0.188 |
| GO:0044839 | cell cycle G2/M phase transition | 266 | -1.618 | 1.15E-04 | 3.86E-03 | 0.238 |
| GO:0022618 | ribonucleoprotein complex assembly | 277 | 1.576 | 1.54E-04 | 5.06E-03 | 0.399 |
| GO:0016236 | macroautophagy | 295 | -1.572 | 1.56E-04 | 5.07E-03 | 0.365 |
| GO:0032886 | regulation of microtubule-based process | 218 | -1.659 | 1.59E-04 | 5.07E-03 | 0.270 |
| GO:0050830 | defense response to Gram-positive bacterium | 101 | -1.909 | 1.59E-04 | 5.07E-03 | 0.400 |
| GO:0034976 | response to endoplasmic reticulum stress | 285 | -1.592 | 1.64E-04 | 5.17E-03 | 0.243 |
| GO:0045132 | meiotic chromosome segregation | 90 | -1.902 | 2.15E-04 | 6.60E-03 | 0.235 |
| GO:0000082 | G1/S transition of mitotic cell cycle | 279 | -1.597 | 2.20E-04 | 6.69E-03 | 0.256 |
| GO:0060969 | negative regulation of gene silencing | 39 | -1.957 | 2.32E-04 | 7.02E-03 | 0.433 |
| GO:0051262 | protein tetramerization | 172 | -1.699 | 2.50E-04 | 7.51E-03 | 0.248 |
| GO:0002474 | antigen processing and presentation of peptide antigen via MHC class I | 96 | -1.780 | 2.52E-04 | 7.53E-03 | 0.333 |
| GO:0032635 | interleukin-6 production | 161 | -1.728 | 2.62E-04 | 7.78E-03 | 0.427 |
| GO:1905268 | negative regulation of chromatin organization | 62 | -1.878 | 2.90E-04 | 8.46E-03 | 0.412 |
| GO:0036498 | IRE1-mediated unfolded protein response | 64 | -1.856 | 3.02E-04 | 8.77E-03 | 0.483 |
| GO:0031167 | rRNA methylation | 27 | 1.942 | 3.04E-04 | 8.77E-03 | 0.440 |
| GO:0043388 | positive regulation of DNA binding | 59 | -1.897 | 3.14E-04 | 9.03E-03 | 0.455 |
| GO:0000470 | maturation of LSU-rRNA | 21 | 1.916 | 4.73E-04 | 1.33E-02 | 0.571 |
| GO:0061640 | cytoskeleton-dependent cytokinesis | 100 | -1.763 | 4.92E-04 | 1.37E-02 | 0.314 |
| GO:0000727 | double-strand break repair via break-induced replication | 11 | -1.888 | 5.04E-04 | 1.40E-02 | 0.500 |
| GO:0008608 | attachment of spindle microtubules to kinetochore | 32 | -1.904 | 5.29E-04 | 1.45E-02 | 0.429 |
| GO:0032715 | negative regulation of interleukin-6 production | 53 | -1.874 | 5.31E-04 | 1.45E-02 | 0.425 |
| GO:0051701 | interaction with host | 209 | -1.602 | 5.64E-04 | 1.52E-02 | 0.297 |
| GO:0032868 | response to insulin | 272 | -1.559 | 6.21E-04 | 1.65E-02 | 0.335 |
| GO:0016052 | carbohydrate catabolic process | 199 | -1.605 | 6.91E-04 | 1.83E-02 | 0.304 |
| GO:0007062 | sister chromatid cohesion | 63 | -1.807 | 7.24E-04 | 1.91E-02 | 0.232 |
| GO:1902969 | mitotic DNA replication | 15 | -1.894 | 7.48E-04 | 1.96E-02 | 0.571 |
| GO:0007033 | vacuole organization | 163 | -1.619 | 7.52E-04 | 1.96E-02 | 0.355 |
| GO:0050764 | regulation of phagocytosis | 98 | -1.748 | 8.13E-04 | 2.09E-02 | 0.468 |
| GO:0002701 | negative regulation of production of molecular mediator of immune response | 35 | -1.884 | 8.17E-04 | 2.09E-02 | 0.269 |
| GO:0042770 | signal transduction in response to DNA damage | 133 | -1.662 | 8.28E-04 | 2.10E-02 | 0.294 |
| GO:0030100 | regulation of endocytosis | 281 | -1.544 | 8.78E-04 | 2.21E-02 | 0.393 |
| GO:0000018 | regulation of DNA recombination | 101 | -1.719 | 8.80E-04 | 2.21E-02 | 0.333 |
| GO:0033260 | nuclear DNA replication | 60 | -1.770 | 1.11E-03 | 2.70E-02 | 0.345 |
| GO:1900026 | positive regulation of substrate adhesion-dependent cell spreading | 37 | -1.846 | 1.22E-03 | 2.97E-02 | 0.552 |
| GO:2001235 | positive regulation of apoptotic signaling pathway | 179 | -1.588 | 1.25E-03 | 3.03E-02 | 0.291 |
| GO:0032465 | regulation of cytokinesis | 89 | -1.726 | 1.40E-03 | 3.33E-02 | 0.279 |
| GO:1902115 | regulation of organelle assembly | 194 | -1.566 | 1.41E-03 | 3.34E-02 | 0.400 |
| GO:0031935 | regulation of chromatin silencing | 37 | -1.831 | 1.44E-03 | 3.38E-02 | 0.367 |
| GO:0031099 | regeneration | 198 | -1.622 | 1.49E-03 | 3.47E-02 | 0.413 |
| GO:0045730 | respiratory burst | 37 | -1.824 | 1.51E-03 | 3.50E-02 | 0.469 |
| GO:0051099 | positive regulation of binding | 179 | -1.583 | 1.53E-03 | 3.53E-02 | 0.276 |
| GO:0018107 | peptidyl-threonine phosphorylation | 126 | -1.639 | 1.57E-03 | 3.60E-02 | 0.264 |
| GO:1904668 | positive regulation of ubiquitin protein ligase activity | 12 | -1.835 | 1.57E-03 | 3.60E-02 | 0.273 |
| GO:0051965 | positive regulation of synapse assembly | 68 | 1.794 | 1.62E-03 | 3.69E-02 | 0.364 |
| GO:0016482 | cytosolic transport | 158 | -1.576 | 1.67E-03 | 3.78E-02 | 0.324 |
| GO:0002385 | mucosal immune response | 36 | -1.837 | 1.71E-03 | 3.85E-02 | 0.542 |
| GO:0071216 | cellular response to biotic stimulus | 236 | -1.529 | 1.72E-03 | 3.86E-02 | 0.385 |
| GO:0044003 | modification by symbiont of host morphology or physiology | 46 | -1.786 | 1.81E-03 | 3.99E-02 | 0.359 |
| GO:0051984 | positive regulation of chromosome segregation | 28 | -1.825 | 1.85E-03 | 4.01E-02 | 0.346 |
| GO:0043044 | ATP-dependent chromatin remodeling | 88 | -1.685 | 1.86E-03 | 4.01E-02 | 0.368 |
| GO:0000083 | regulation of transcription involved in G1/S transition of mitotic cell cycle | 29 | -1.827 | 1.88E-03 | 4.05E-02 | 0.280 |
| GO:0051053 | negative regulation of DNA metabolic process | 156 | -1.588 | 1.94E-03 | 4.16E-02 | 0.277 |
| GO:0031330 | negative regulation of cellular catabolic process | 264 | -1.495 | 2.09E-03 | 4.39E-02 | 0.336 |
| GO:0048771 | tissue remodeling | 179 | -1.613 | 2.11E-03 | 4.39E-02 | 0.367 |
| GO:0002573 | myeloid leukocyte differentiation | 204 | -1.541 | 2.16E-03 | 4.46E-02 | 0.344 |
| GO:0003014 | renal system process | 120 | -1.691 | 2.17E-03 | 4.46E-02 | 0.232 |
| GO:0030262 | apoptotic nuclear changes | 35 | -1.802 | 2.21E-03 | 4.50E-02 | 0.483 |
| GO:0046653 | tetrahydrofolate metabolic process | 19 | -1.829 | 2.23E-03 | 4.53E-02 | 0.438 |
| GO:0050777 | negative regulation of immune response | 150 | -1.596 | 2.41E-03 | 4.81E-02 | 0.248 |

**Legend**

**Size** Number of genes defining the biological process in the GO database
**NES** Normalize Enrichment Score, indicating the level of enrichment and its directionality
**pvalue** Empirical statistical significance value for enrichment (n=1e6 permutations)
**p.adjust** FDR-corrected p-value
**ratio*** Ratio between #core BP enriched genes and #total genes in biological process

*As described in https://www.gsea-msigdb.org/gsea/doc/GSEAUserGuideFrame.html

**Supplementary Table 3: List of overlapping biologial processes between COVID19 and abatacept exposures**

| ID | Description | Size | NES_COVID | pCOVID | ratio_COVID | NES_ABT | pABT | ratio_ABT |
|---|---|---|---|---|---|---|---|---|
| GO:0000070 | mitotic sister chromatid segregation | 151 | 2.028 | 1.08E-04 | 0.273 | -2.243 | 1.28E-04 | 0.382 |
| GO:0000184 | nuclear-transcribed mRNA catabolic process, nonsense-mediated decay | 120 | -1.808 | 1.48E-03 | 0.576 | 2.663 | 1.28E-04 | 0.563 |
| GO:0001510 | RNA methylation | 81 | -1.737 | 1.48E-02 | 0.279 | 1.953 | 8.25E-04 | 0.359 |
| GO:0002385 | mucosal immune response | 36 | -1.784 | 4.68E-02 | 0.444 | -1.837 | 3.85E-02 | 0.542 |
| GO:0002440 | production of molecular mediator of immune response | 286 | 1.918 | 1.08E-04 | 0.442 | -2.758 | 1.28E-04 | 0.216 |
| GO:0002455 | humoral immune response mediated by circulating immunoglobulin | 150 | 3.043 | 1.08E-04 | 0.739 | -3.253 | 1.28E-04 | 0.612 |
| GO:0002474 | antigen processing and presentation of peptide antigen via MHC class I | 96 | 2.509 | 1.08E-04 | 0.505 | -1.780 | 7.53E-03 | 0.333 |
| GO:0002526 | acute inflammatory response | 220 | 2.600 | 1.08E-04 | 0.558 | -3.092 | 1.28E-04 | 0.379 |
| GO:0002573 | myeloid leukocyte differentiation | 204 | 1.656 | 4.97E-03 | 0.497 | -1.541 | 4.46E-02 | 0.344 |
| GO:0006364 | rRNA processing | 214 | -1.717 | 1.13E-03 | 0.420 | 2.183 | 1.28E-04 | 0.419 |
| GO:0006614 | SRP-dependent cotranslational protein targeting to membrane | 105 | -2.113 | 1.09E-04 | 0.611 | 2.746 | 1.28E-04 | 0.704 |
| GO:0006890 | retrograde vesicle-mediated transport, Golgi to ER | 86 | 1.818 | 4.72E-03 | 0.455 | -1.947 | 1.13E-03 | 0.418 |
| GO:0007033 | vacuole organization | 163 | 2.177 | 1.08E-04 | 0.459 | -1.619 | 1.96E-02 | 0.355 |
| GO:0008037 | cell recognition | 215 | 2.040 | 1.76E-04 | 0.535 | -2.611 | 1.28E-04 | 0.328 |
| GO:0009451 | RNA modification | 163 | -1.523 | 3.29E-02 | 0.264 | 1.829 | 2.97E-04 | 0.294 |
| GO:0010324 | membrane invagination | 135 | 2.441 | 1.08E-04 | 0.701 | -2.837 | 1.28E-04 | 0.418 |
| GO:0016052 | carbohydrate catabolic process | 199 | 2.129 | 1.08E-04 | 0.393 | -1.605 | 1.83E-02 | 0.304 |
| GO:0016236 | macroautophagy | 295 | 2.152 | 1.08E-04 | 0.432 | -1.572 | 5.07E-03 | 0.365 |
| GO:0019083 | viral transcription | 177 | -1.732 | 1.32E-03 | 0.465 | 2.439 | 1.28E-04 | 0.494 |
| GO:0019882 | antigen processing and presentation | 226 | 2.494 | 1.08E-04 | 0.475 | -1.972 | 1.28E-04 | 0.369 |
| GO:0022618 | ribonucleoprotein complex assembly | 277 | -1.471 | 2.36E-02 | 0.294 | 1.576 | 5.06E-03 | 0.399 |
| GO:0030100 | regulation of endocytosis | 281 | 1.998 | 1.08E-04 | 0.477 | -1.544 | 2.21E-02 | 0.393 |
| GO:0032635 | interleukin-6 production | 161 | 1.859 | 4.99E-04 | 0.505 | -1.728 | 7.78E-03 | 0.427 |
| GO:0032868 | response to insulin | 272 | 1.641 | 2.93E-03 | 0.402 | -1.559 | 1.65E-02 | 0.335 |
| GO:0034976 | response to endoplasmic reticulum stress | 285 | 1.564 | 5.95E-03 | 0.416 | -1.592 | 5.17E-03 | 0.243 |
| GO:0038093 | Fc receptor signaling pathway | 241 | 2.466 | 1.08E-04 | 0.524 | -3.021 | 1.28E-04 | 0.371 |
| GO:0038094 | Fc-gamma receptor signaling pathway | 142 | 2.607 | 1.08E-04 | 0.642 | -3.182 | 1.28E-04 | 0.531 |
| GO:0044839 | cell cycle G2/M phase transition | 266 | 1.740 | 3.08E-04 | 0.379 | -1.618 | 3.86E-03 | 0.238 |
| GO:0045132 | meiotic chromosome segregation | 90 | 1.738 | 2.60E-02 | 0.234 | -1.902 | 6.60E-03 | 0.235 |
| GO:0045730 | respiratory burst | 37 | 1.930 | 9.65E-03 | 0.552 | -1.824 | 3.50E-02 | 0.469 |
| GO:0048771 | tissue remodeling | 179 | 1.894 | 5.85E-04 | 0.505 | -1.613 | 4.39E-02 | 0.367 |
| GO:0050764 | regulation of phagocytosis | 98 | 1.880 | 1.87E-03 | 0.547 | -1.748 | 2.09E-02 | 0.468 |
| GO:0050864 | regulation of B cell activation | 184 | 1.949 | 1.08E-04 | 0.500 | -2.765 | 1.28E-04 | 0.326 |
| GO:0051225 | spindle assembly | 108 | 1.833 | 1.87E-03 | 0.284 | -2.041 | 1.28E-04 | 0.443 |
| GO:0051262 | protein tetramerization | 172 | 1.812 | 1.12E-03 | 0.311 | -1.699 | 7.51E-03 | 0.248 |
| GO:0051304 | chromosome separation | 90 | 1.965 | 5.85E-04 | 0.267 | -2.215 | 1.28E-04 | 0.297 |
| GO:0051321 | meiotic cell cycle | 249 | 1.596 | 1.30E-02 | 0.275 | -1.721 | 3.60E-03 | 0.188 |
| GO:0051701 | interaction with host | 209 | 1.707 | 1.64E-03 | 0.416 | -1.602 | 1.52E-02 | 0.297 |
| GO:0051965 | positive regulation of synapse assembly | 68 | -1.788 | 4.36E-02 | 0.650 | 1.794 | 3.69E-02 | 0.364 |
| GO:0051984 | positive regulation of chromosome segregation | 28 | 1.864 | 2.02E-02 | 0.240 | -1.825 | 4.01E-02 | 0.346 |
| GO:0061640 | cytoskeleton-dependent cytokinesis | 100 | 1.688 | 2.03E-02 | 0.417 | -1.763 | 1.37E-02 | 0.314 |
| GO:0070125 | mitochondrial translational elongation | 88 | 1.613 | 3.33E-02 | 0.481 | 1.892 | 1.60E-03 | 0.322 |
| GO:0071216 | cellular response to biotic stimulus | 236 | 1.756 | 8.42E-04 | 0.421 | -1.529 | 3.86E-02 | 0.385 |
| GO:0140014 | mitotic nuclear division | 264 | 1.766 | 1.76E-04 | 0.227 | -2.222 | 1.28E-04 | 0.394 |
| GO:1901216 | positive regulation of neuron death | 94 | 1.742 | 1.08E-02 | 0.533 | -1.930 | 1.54E-03 | 0.382 |
| GO:1901988 | negative regulation of cell cycle phase transition | 267 | 1.757 | 3.65E-04 | 0.432 | -1.811 | 1.28E-04 | 0.295 |
| GO:1902850 | microtubule cytoskeleton organization involved in mitosis | 131 | 1.891 | 4.99E-04 | 0.330 | -2.213 | 1.28E-04 | 0.461 |
| GO:1904668 | positive regulation of ubiquitin protein ligase activity | 12 | 1.889 | 2.05E-02 | 0.545 | -1.835 | 3.60E-02 | 0.273 |
| GO:1905819 | negative regulation of chromosome separation | 40 | 1.704 | 4.89E-02 | 0.324 | -2.197 | 1.92E-04 | 0.382 |

**Legend**

| | |
|---|---|
| **ID** | GO term ID |
| **Size** | Number of genes defining the biological process in the GO database |
| **NES COVID** | Normalize Enrichment Score for the COVID19 dataset |
| **pCOVID** | FDR-corrected p-value |
| **ratio COVID** | Ratio between #enriched genes and #total genes in biological process |
| **NES RA Abatac** | Normalize Enrichment Score for the abatacept dataset |
| **pABT** | FDR-corrected p-value |
| **ratio ABT** | Ratio between #enriched genes and #total genes in biological process |